\documentclass[transmag]{IEEEtran}
\usepackage{latexsym}
\usepackage{graphicx}
\usepackage{amsfonts,amssymb,amsmath}
\usepackage{hyperref}
\usepackage[T1]{fontenc}
\usepackage{setspace}
\usepackage{color}
\usepackage{babel}
\usepackage{float}
\usepackage[center]{caption}
\usepackage{textcomp}
\usepackage{amsmath}
\usepackage{amssymb}
\usepackage{graphicx}
\usepackage{algorithm}
\usepackage{algpseudocode}
\usepackage{cite}
\def\BibTeX{{\rm B\kern-.05em{\sc i\kern-.025em b}\kern-.08em T\kern-.1667em\lower.7ex\hbox{E}\kern-.125emX}}
\begin{document}

\title{HARQ in Full-Duplex Relay-Assisted Transmissions for URLLC}

\author{Fatima Ezzahra~Airod, Houda~Chafnaji, and Halim~Yanikomeroglu%
\thanks{Fatima Ezzahra~Airod is with the Department of Communication Systems,
INPT, Rabat, Morocco, e-mail: \protect\href{http://airod@inpt.ac.ma}{airod@inpt.ac.ma}.%
}%
\thanks{Houda~Chafnaji is with the Department of Communication Systems, INPT,
Rabat, Morocco, e-mail: \protect\href{http://chafnaji@inpt.ac.ma}{chafnaji@inpt.ac.ma}.%
}%
\thanks{Halim~Yanikomeroglu is with the Department of Systems and Computer
Engineering, Carleton University, Ottawa, Canada, e-mail: \protect\href{http:// halim@sce.carleton.ca}{ halim@sce.carleton.ca}.%
}}

\IEEEtitleabstractindextext{\begin{abstract}The Release 16 completion unlocks the road to an exciting phase pertain to the sixth generation (6G) era. Meanwhile, to sustain far-reaching applications with unprecedented challenges in terms of latency and reliability, much interest is already getting intensified toward physical layer specifications of 6G. In support of this vision, this work exhibits the forward-looking perception of full-duplex (FD) cooperative relaying in support of upcoming generations and adopts as a mean concern the critical contribution of hybrid automatic repeat request (HARQ) mechanism to ultra-reliable and low-latency communication (URLLC).  Indeed, the HARQ roundtrip time (RTT) is known to include basic physical delays that may cause the HARQ abandonment for the 1 ms latency use case of URLLC. Taking up these challenges, this article proposes a hybrid FD amplify-and-forward (AF)-selective decode-and-forward (SDF) relay-based system for URLLC. Over this build system, two HARQ procedures within which the HARQ RTT is shortened, are suggested to face latency and reliability issues, namely, the conventional and the enhanced HARQ procedures. We develop then an analytical framework of this relay based HARQ system within its different procedures. Finally, using Monte-Carlo simulations, we confirm the theoretical results and compare the proposed relay-assisted HARQ procedures to the source-to-destination (S2D) HARQ-based system where no relay assists the communication between the source and the destination.\end{abstract}


}

\maketitle

\section{INTRODUCTION}

\IEEEPARstart{U}{ltra}-reliable and low-latency communication (URLLC) is one of fifth generation (5G) driven use cases, for which the third partnership project (3GPP) and the International Mobile Telecommunications 2020 and beyond (IMT-2020) specified stringent latency and reliability requirements to fulfill. The 3GPP Rel. 14 has defined these requirements to at most 1 ms latency with an outage probability of less than $10^{-5}$ for a 32 byte packet \cite{rel14,Pacovi2018}. In fact, the set of URLLC services with these demanding requirements in the real world, is inappropriate with the best effort manner of long term evolution (LTE) \cite{strodthoff}. To this end, the 5G New Radio (NR) has been standardized with a set of specifications in Rel. 15, which incorporates basic URLLC features. Indeed, the NR covers implicitly the air interface latency aspect by the considered 5G numerology, however, it is not developed enough to fully achieve the required latencies. In fact, still the retransmission process across the air-interface unclear, in particular, the hybrid automatic repeat request (HARQ) roundtrip time (RTT) which has to be shortened to fit in the assumed 1 ms latency budget. Otherwise, the contribution of HARQ to URLLC is too critical and might be then, limited \cite{srveyURLLC}. However, recently, as part of Rel. 16, different areas of improvements have been specified for the NR URLLC foundation. One of these enhancements focused on the uplink control information (UCI), eventually, the HARQ feedback. In fact, the main purpose is to enable the NR support for applications with latency requirements in the range of 0.5 to 1 ms and where network reliability is critical \cite{rel16_stg2}. Along with this, other technological capabilities have been included to enable real-time situations, namely, the sidelink concept specifically for cellular-vehicle-to-everything (C-V2X) communications and integrated access backhaul (IAB) that points out mainly the relay concept upon 3GPP NR. Yet, several extensions will be supported by a Rel. 17 work item in order to bring several  IAB improvements in terms of spectral efficiency (SE), latency, and performance. In other words Rel. 17 will rejuvenate the interest toward full-duplex (FD) transmission mode by proposing a multiplexing option for transmissions between the backhaul and access links, referred as, IAB-node full-duplex \cite{rel17}.  These changes mark undeniably the dawn of 6G era toward which the rich theory of FD cooperative communications is getting unleashed to aid the full sustainability of the next generation driving use cases with stringent requirements in terms of latency and reliability \cite{rel16_2}. 

Cooperative relaying and FD radio communication have been extensively identified as a promising network technology to improve wireless communications reliability. In other words, as the capacity improvement is promoted by the SE improvement, the adoption of FD communication at the relay node is more advantageous and efficient. Meanwhile, thanks to recent advances noted in antenna technology and signal processing techniques, it is practical to use the FD communication mode on cooperative relaying systems even with the presence of a significant loop interference \cite{rsiCancel1,FDcancel,FD2}. Furthermore, in the perspective of a low latency, FD relaying mode allows fast device-to-device discovery, and hence, contributes on the delay reduction. Accordingly, all these practical growth have  incited authors to adopt FD communications in their researches, thus, get rid of the spectral inefficiency caused by half-duplex (HD) relaying mode \cite{FD1,c3}.

Likewise, cooperative relay communications have been always the focus of several works in the literature. Nevertheless, to the best of our knowledge their full-fledged implementation has never occurred. Indeed, the relaying concept has been introduced in 3GPP Rel. 10 as a part of LTE-Advanced, and currently, it get covered again in some study item (SI) proposals for the NR 3GPP rel. 17, specifically, pertain to SL and IAB concepts \cite{rel17 , rel17_2}. In fact,  relay nodes were formerly, used for the coverage extension purpose, i.e., the in-coverage user equipment (UE) assists the communication between an evolved node B (eNB) and an out-of-coverage UE \cite{c1}, there is no combining of signals. Yet recently, cooperative relaying recurs intensively in the literature with great interest for competitive edge of wireless networks while targeting far-reaching use cases \cite{c2,c3,c4,c5,c6,c7,c8,c9,c10}. In support of this vision and combined with HARQ, cooperative relaying can substantially improve the system reliability and thus aid upgrading the new functions of Rel. 16 \cite{jaz62}. Packet retransmission in a cooperative network involves a source terminal, a destination terminal and at
least one relay terminal, and it consists on recovering corrupted
data packets from neighboring nodes. In other words, if the destination
fails to decode the received signal, it requests the retransmission
of the packet from neighboring nodes, i.e., relays, rather than asking
the source node. This provides the system with a spatial diversity
gain since the receiver obtains multiple copies of the same packet
on channels experiencing independent fading. Therefore, link reliability
as well as overall throughput can be improved. 

However, when adopting cooperative relay communications, a drastic change will affect the communication latency which may bring further challenges to establish a successful quasi-real time radio communication. Generally, the supplementary latency due to relaying depends mainly on the investigated relaying technique. These techniques can be classified into regenerative and non-regenerative classes, among which we distinct respectively, the most commonly used selective decode and forward (SDF) and amplify-and-forward (AF) \cite{SDF, Lan}. In order to recognize the impact of the relay processing delay on the system latency, an exhaustive comparison between AF and SDF techniques under the framework of full-duplex (FD) relaying has been conducted in \cite{airod}. Simulations results proved that the SDF block-based transmission scheme is no longer practical to adopt for FD signal-combining mode, where direct and relay links are combined at the receiver side, especially, with the extra latency induced due to the complexity of encoding and decoding algorithms. Still the AF scheme represents better choice in terms of outage performance and latency.

In the literature, HARQ and relaying have been studied extensively.
In \cite{harq_litt1}, authors have investigated the performance of
relay networks in the presence of HARQ feedback and adaptive power
allocation. The throughput and the outage probability of different
HARQ protocols were studied for independent and spatially-correlated
fading channels. In \cite{han63}, with the joint optimization of
transmission power and rate, authors have adopted and have compared between many
packet retransmissions schemes to find out the most appropriate one
that yields to the best spectrum efficiency performance over multi-hop
relay networks. In \cite{Harq_litt2,harq_litt3}, new types of ARQ
based on a selective and opportunistic AF relaying method have been
developed. Through simulations, authors showed that the proposed
ARQ schemes are more effective for throughput enhancements, and can
provide cell-edge users almost three times the throughput gain in
comparison with ARQ with no relay-assisted forwarding. However, in
the different mentioned studies, authors have applied some features
to the existing retransmission mechanism with a specific purpose:
Increasing the system reliability. However, this increase in reliability
comes at the cost of higher latency at the system air-interface. In
\cite{harq_litt4}, authors have investigated diamond relay systems
with HARQ protocol under delay constraints and have employed the effective
capacity as a performance metric. However, the studied system has
assumed a disconnected source and destination, which means that there
is no direct link between the two nodes. In \cite{harq_litt5}, authors have investigated a performance evaluation of HARQ-assisted hybrid satellite-terrestrial relay systems. They considered as a key performance metric, the delay limited throughput, then provided a convenient approach that includes different HARQ schemes, different relay protocols, and different key parameters. Authors in \cite{harq_litt6}, conduct a generic analysis of HARQ with code combining (HARQ-CC) over double Rayleigh channels, which they consider as a typical framework that can be applied to scenarios pertains to urban vehicle-to-vehicle communication systems, amplify-and-forward relaying, as well as keyhole channels.

To the best of our knowledge, none of the cited works have build up or evaluated packet retransmission mechanisms relative to latency purpose under the framework of full-fledged implemented FD cooperative systems with non-negligible direct link. Motivated by this fact, and  inspired by the comparative study results between
FD-AF and FD-SDF relaying schemes presented in \cite{airod}, we investigate a hybrid FD AF-SDF relay-based system for URLLC. The proposed hybrid relay assists the communication using the low latency FD-AF relaying scheme during the initial transmission. If the destination fails in decoding the received packet, the hybrid relay retransmits the same message but processed using SDF relaying. 
 
The main contributions of this paper can be summarized as follows:
\begin{enumerate}
\item From cross-layer perspective, this article points out the conflicting
contribution of the HARQ mechanism for URLLC services. In this work, our main concern is to improve the system reliability while respecting the 1 ms latency budget requested by URLLC. Thereby, we propose the use of FD-AF during the first transmission which highly improves the signal quality at the destination side and thereby reduces the retransmission probability.
 
\item For the adopted hybrid relay-assisted scheme, we propose two HARQ-based
procedures aiming at keeping a trade-off between an extreme reliability
and a low latency. Both of procedures were not considered before under
the framework of FD cooperative systems with non-negligible direct
link. We first propose the conventional HARQ procedure for a relay-based
system where the retransmission of erroneous packets is taken over
by the relay node. Then we focus on latency and reliability issues.
So as to face the latency issue, we assume a reduced HARQ roundtrip
time as well as a consecutive transmission time intervals (TTIs) assignment
using multiple HARQ processes. The enhanced HARQ procedure is then
proposed to deal with the reliability issue in the case of bad source-to-relay
link. 
\item We evaluate the outage behavior of this relay based HARQ system within
its different procedures. Note that the analytical development for the first transmission is similar to \cite{airod}. Thereby, this work's main technical contribution in term of outage probability concerns the second transmission where the multiple copies of the signal (from the source and the hybrid FD AF-SDF relay) are combined coherently. 
\item  For the purpose of reducing the latency, we assume a short TTI of 0.125 ms obtained at subcarrier spacing of 120 kHz. Then to evaluate the performance of our two proposed
HARQ procedures, we investigate extensive simulation performance analysis where we
include as a reference, the source-to-destination (S2D) HARQ-based
system where no relay assists the communication between the source
and the destination, the AF relaying scheme as the low latency relay
processing scheme, and the SDF relaying scheme that requires complex
encoding and decoding algorithms. 
\end{enumerate}
The rest of the paper is organized as follows: Section \ref{sec:Relay-System-model}
presents the system model. In Section \ref{sec:HARQ procedures},
HARQ procedures are elaborated in detail. Outage probability is derived
in Section \ref{sec:Outage-Probability}. In Section \ref{sec:Numerical-Results},
numerical results are shown and discussed. The paper is concluded
in Section \ref{sec:Conclusion}.

\section{\label{sec:Relay-System-model}System Model}

We consider a single relay cooperative system, where one relay $(\mathrm{R})$
assists the communication between a source $(\mathrm{S})$ and a destination
$(\mathrm{D})$. Since the relay operates in FD mode, we take into
account the residual self-interference (RSI) generated from relay's
input to relay's output. In this work, we assume the direct link between
the source and the destination nodes is non-negligible and all channels
$h_{\mathrm{AB}}$, with $\mathrm{A}\in\left\{ \mathrm{S},\,\mathrm{R}\right\} $
and $\mathrm{B}\in\left\{ \mathrm{R},\,\mathrm{D}\right\} $
are i.i.d zero mean circularly symmetric complex Gaussian $\sim\mathcal{CN}(0,\sigma_{\mathrm{AB}}^{2})$.  Moreover, we assume a limited channel
state information (CSI) at the source node, i.e., the source transmitter is only aware
of the processing delay at the relay, and we suppose perfect CSI at the relay and the
destination receivers.

The proposed HARQ-based system consists of three phases. Phase I represents
the packet initial transmission where the source sends the data packet
to the destination by the mean of a single relay, using FD-AF relaying.
In phase II, the destination checks the correctness of the received
packet and transmits either a positive or negative acknowledgment
(ACK/NACK) in order to convey the success or failure of data decoding.
Phase III is the packet retransmission phase activated if the destination
responds with a NACK. In fact, if the destination fails in decoding
a message, it stores the received signal and responds with a NACK.
During the retransmission phase (Phase III), the relay resends the
same message but processed using SDF relaying instead of AF relaying.
At the destination, the phase I received signal is combined with phase
III received signal using maximum ratio combining (MRC). In this work,
we suppose perfect packet error detection and assume that the NACK
feedback message is error free. Hereafter, we investigate the communication
model explicitly.

During Phase I, the source first broadcasts its signal $x_{\mathrm{s}}$,
to both the relay and destination nodes. Due to the processing delay
at the relay, the destination node will receive the source node transmitted
signal $x_{\mathrm{s}}$ at different channel uses. In order to alleviate
the inter-symbol interference (ISI) caused by the delayed signal,
a cyclic-prefix (CP) transmission can be used at the source side \cite{airod}.
Therefore, before transmitting, the source node appends a CP of length
$\tau$ to the transmitted signal, with $\tau$ represents the AF
processing delay at the relay. The FD relay forwards then, the amplified
signal, i.e., $x_{\mathrm{R}},$ to the destination upon the same
time slot and starts the SDF processing %
\footnote{The proposed FD relay decodes the received signal then reencodes it
in order to retransmit it to the destination during Phase III. Note
that, if the relay decoding outcome is erroneous, the relay keeps
silent during the retransmission phase (Phase III).%
}.

The received signal at the destination side, during Phase I, and after
the CP removal, is expressed as

\begin{align}
y_{\mathrm{D}}^{I}(t) & =\underset{\mathrm{Direct\,+\, Relayed\, signal}}{\underbrace{\sqrt{P_{\mathrm{s}}}h_{\mathrm{SD}}x_{\mathrm{s}}(t)+\beta\sqrt{P_{\mathrm{s}}}h_{\mathrm{RD}}h_{\mathrm{SR}}x_{\mathrm{s}}\left[(t-\tau)\,\mathrm{mod}\, T\right]}}\label{eq:2-1}\\
 & +\underset{\mathrm{RSI}}{\underbrace{\beta h_{\mathrm{RD}}h_{\mathrm{RR}}x_{\mathrm{R}}(t-\tau)}}+\underset{\mathrm{Noise}}{\underbrace{\beta h_{\mathrm{RD}}n_{\mathrm{R}}(t-\tau)+n_{\mathrm{D}}(t)}},\nonumber 
\end{align}
with $\beta=\sqrt{\frac{P_{\mathrm{R}}}{P_{\mathrm{s}}|h_{\mathrm{SR}}|^{2}+P_{\mathrm{R}}\sigma_{\mathrm{RR}}^{2}+N_{R}}}$
is the amplification constant factor \textcolor{black}{chosen to satisfy
the total power constraint at the relay} \cite{amplif}, $P_{\mathrm{s}}$
and $P_{\mathrm{R}}$ denote, respectively, the transmit power at
the source and the relay, and $h_{\mathrm{RR}}x_{\mathrm{R}}(t)$
is the RSI after undergoing any
cancellation techniques and practical isolation at the relay \cite{rsiCancel1,rsiCancel2}.
$n_{\mathrm{R}}\sim\mathcal{CN}(0,N_{R})$ and $n_{\mathrm{D}}\sim\mathcal{CN}(0,N_{\mathrm{D}})$
denote, respectively, a zero-mean complex additive white Gaussian
noise at the relay and the destination. Without loss of generality,
we assume that the source transmitted signals are independent $\mathbb{{E}}\left[x_{\mathrm{s}}(i)x_{\mathrm{s}}^{\star}(i')\right]=\delta_{i,i'}$.

In this paper, we assume that all channel gains change independently
from one block to another and remain constant during one block of
$T+\tau$ channel uses, with $T$ represents the number of transmitted
codewords. Hence, we can rewrite (\ref{eq:2-1}) in vector form to
jointly take into account the $T+\tau$ received signal as \cite{turbo}

\begin{equation}
\mathbf{y_{\mathrm{D}}^{I}}=\boldsymbol{{\mathcal{H}}}\mathbf{x}_{\mathrm{s}}+\beta h_{\mathrm{RD}}h_{\mathrm{RR}}\mathbf{x}_{\mathrm{R}}+\mathbf{n},
\end{equation}
where $\mathbf{y_{\mathrm{D}}^{I}}=\left[y_{D}\left(0\right),...,y_{D}\left(T-1\right)\right]^{\top}\in\mathbb{C}^{T\times1}$,
$\mathbf{x}_{\mathrm{s}}=\left[x_{\mathrm{s}}\left(0\right),...,x_{\mathrm{s}}\left(T-1\right)\right]^{\top}\in\mathbb{C}^{T\times1}$,
$\mathbf{x}_{\mathrm{R}}=\left[\tilde{x}_{\mathrm{R}}\left(0\right),...,\tilde{x}_{\mathrm{R}}\left(T-1\right)\right]^{\top}\in\mathbb{C}^{T\times1}$
with $\tilde{x}_{\mathrm{R}}(i)=x_{\mathrm{R}}(t-\tau)$, $\mathbf{n}=\left[n\left(0\right),...,n\left(T-1\right)\right]^{\top}\in\mathbb{C}^{T\times1}$
with $n(t)=\beta h_{\mathrm{RD}}n_{\mathrm{R}}(t-\tau)+n_{\mathrm{D}}(t)$,
and $\boldsymbol{{\mathcal{H}}}\in\mathbb{C}^{T\times T}$ is a circulant
matrix whose first column matrix is $\left[\sqrt{P_{\mathrm{s}}}h_{\mathrm{SD}},\boldsymbol{{0}}_{1\times\tau-2},\beta\sqrt{P_{\mathrm{s}}}h_{\mathrm{RD}}h_{\mathrm{SR}},\boldsymbol{{0}}_{1\times T-\tau}\right]^{\top}$. 

Note that the circulant matrix $\boldsymbol{{\mathcal{H}}}$, can
be decomposed as

\begin{equation}
\boldsymbol{{\mathcal{H}}}=\mathbf{U}_{T}^{H}\boldsymbol{{\Lambda}}\mathbf{U}_{T},
\end{equation}
where $\mathbf{U}_{T}$\footnote{Generally, for $\mathbf{x}\in\mathbb{C}^{T\times1}$, $\mathbf{x}_{f}$
denotes the discrete Fourier transform (DFT) of $\mathbf{x}$, i.e. $\mathbf{x}_{f}=\mathbf{U}_{T}\mathbf{x}$.}  is a unitary $T\times T$ matrix whose $\left(m,n\right)th$ element is $\left(\mathbf{U}_{T}\right)_{m,n}=\frac{1}{\sqrt{T}}e^{-j(2\pi mn/T)}$ , $j=\sqrt{-1}$
  and $\boldsymbol{{\Lambda}}$ is a diagonal matrix whose $\left(i,i\right)$-th
element is

\begin{equation}
\lambda_{i}=\sqrt{P_{\mathrm{s}}}h_{\mathrm{SD}}+\beta\sqrt{P_{\mathrm{s}}}h_{\mathrm{RD}}h_{\mathrm{SR}}e^{-j\left(2\pi i\frac{\tau}{T}\right)}.
\end{equation}

The signal $\mathbf{y^{I}}_{\mathrm{D}}$ can be therefore represented
in the frequency domain as

\begin{equation}
\mathbf{y_{\mathrm{D}_{f}}^{I}}=\boldsymbol{{\Lambda}}\mathbf{x}_{\mathrm{s}_{f}}+\beta h_{\mathrm{RD}}h_{\mathrm{RR}}\mathbf{x}_{R_{f}}+\mathbf{n}_{f}.
\end{equation}

At the destination side, the instantaneous end-to-end equivalent signal-to-interference
and noise ratio (SINR), during Phase I, is expressed as

\begin{equation}
\begin{aligned}\gamma_{i}^{I}= & \frac{\lambda_{i}\lambda_{i}^{H}}{\beta^{2}|h_{\mathrm{RD}}|^{2}(P_{\mathrm{R}}\sigma_{\mathrm{RR}}^{2}+N_{R})+N_{\mathrm{D}}}\\
= & \rho_{I}+2\left|\mu_{I}\right|\cos\left(2\pi i\frac{\tau}{T}+\theta\right),
\end{aligned}
\label{eq:phase1}
\end{equation}
with {$\rho_{I}=\frac{P_{\mathrm{S}}|h_{\mathrm{\mathrm{S}D}}|^{2}+\beta^{2}P_{\mathrm{S}}|h_{\mathrm{SR}}|^{2}|h_{\mathrm{RD}}|^{2}}{\beta^{2}|h_{\mathrm{RD}}|^{2}(P_{\mathrm{R}}\sigma_{\mathrm{RR}}^{2}+N_{R})+N_{\mathrm{D}}}$,}\\
{$\mu_{I}=\frac{\beta P_{\mathrm{s}}h_{\mathrm{SD}}h_{\mathrm{RD}}h_{\mathrm{SR}}}{\beta^{2}|h_{\mathrm{RD}}|^{2}(P_{\mathrm{R}}\sigma_{\mathrm{RR}}^{2}+N_{R})+N_{\mathrm{D}}}$}, and {$\theta=\textrm{angle}\left(h\mathrm{_{SD}},h_{\mathrm{RD}}^{*}\right)$.} 

During Phase II, the destination verifies the correctness of the received
data packet. If the decoding outcome is correct, the communication
moves on to Phase I and starts the transmission of a new information
block. When the destination fails in decoding a message, it stores
the received signal and the packet retransmission phase is activated.
The retransmission strategy depends on the considered HARQ procedure.
In general, the received signal at the destination, during Phase III,
can be expressed as

\begin{equation}
y_{\mathrm{D}}^{III}(t)=\left(\underset{\mathrm{Retransmitted\, signal}}{\underbrace{\sqrt{P_{\mathrm{X}}}h_{\mathrm{\mathrm{X}D}}^{III}x_{\mathrm{s}}(t)}}+n_{\mathrm{D}}^{III}(t)\right)\times a,\label{eq:combinDest-2-1}
\end{equation}
 with $a=\begin{cases}
1 & \textrm{if Phase III is activated}\\
0 & \textrm{Otherwise}
\end{cases}$ \\and $\textrm{X}\in\left\{ \textrm{S,R}\right\} $ is the retransmitter
node. 

After combining the resulted copies of the same data packet at the
destination using MRC technique, the instantaneous end-to-end SINR,
at the destination and during Phase III is given by \textcolor{red}{{} }

\begin{align}
\gamma_{i}^{III} & =\frac{P_{\mathrm{S}}|h_{\mathrm{\mathrm{S}D}}|^{2}+\beta^{2}P_{\mathrm{S}}|h_{\mathrm{SR}}|^{2}|h_{\mathrm{RD}}|^{2}}{\beta^{2}|h_{\mathrm{RD}}|^{2}(P_{\mathrm{R}}\sigma_{\mathrm{RR}}^{2}+N_{R})+N_{\mathrm{D}}}+\label{eq:}\\
 & \frac{2\left|\beta P_{\mathrm{s}}h_{\mathrm{SD}}h_{\mathrm{RD}}h_{\mathrm{SR}}\right|cos\left(2\pi i\frac{\tau}{T}+\theta\right)}{\beta^{2}|h_{\mathrm{RD}}|^{2}(P_{\mathrm{R}}\sigma_{\mathrm{RR}}^{2}+N_{R})+N_{\mathrm{D}}}+\frac{P_{\textrm{X}}|h_{\mathrm{XD}}^{III}|^{2}}{N_{\mathrm{D}}}\times a.\nonumber 
\end{align}

\section{\label{sec:HARQ procedures}Relay-based HARQ Procedures}

The goal of this section is to propose HARQ procedures for a relay
assisted system in order to improve the latency performance. For that
purpose, we first present the proposed conventional HARQ procedure
for a relay-based system and focus on latency and reliability issues.
To face the latency issue, we assume a reduced HARQ RTT \cite{RTT1, RTT2, RTT3} as well
as a consecutive TTI assignment using multiple HARQ processes. The
enhanced HARQ procedure is then proposed to deal with the reliability
issue in the case of a bad source-to-relay link.

\subsection{Relay-based Conventional HARQ Procedure}

\begin{figure}[tbh]
\begin{centering}
\includegraphics[scale=0.55]{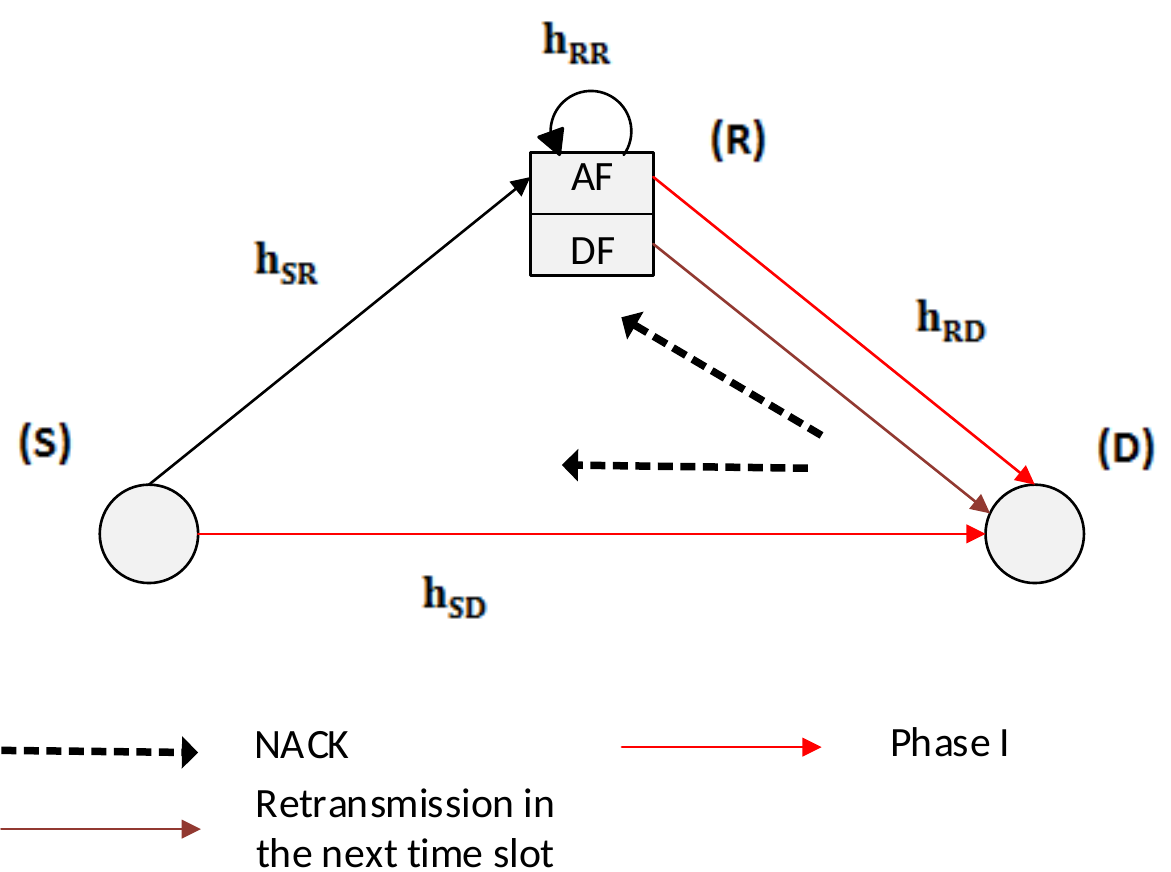}
\par\end{centering}

\protect\caption{\label{fig:System-model-of2}Conventional HARQ procedure diagram. }
\end{figure}

In the conventional system, the destination checks the correctness
of the Phase I received packet and transmits either a NACK or ACK
in order to convey the failure or success of data decoding. When the
destination fails in decoding a message, it stores the received signal
and responds with NACK. The NACK is broadcasted to both relay and
source nodes, as represented in Fig. \ref{fig:System-model-of2}. While
the source keeps silent, the relay retransmits the SDF processed packet.
The received signal at the destination, during Phase III, is expressed
as in (\ref{eq:combinDest-2-1}) with $\textrm{X}=\textrm{R}$.

\subsubsection{Latency Issue}

\begin{figure}[tbh]
\begin{centering}
\includegraphics[scale=0.55]{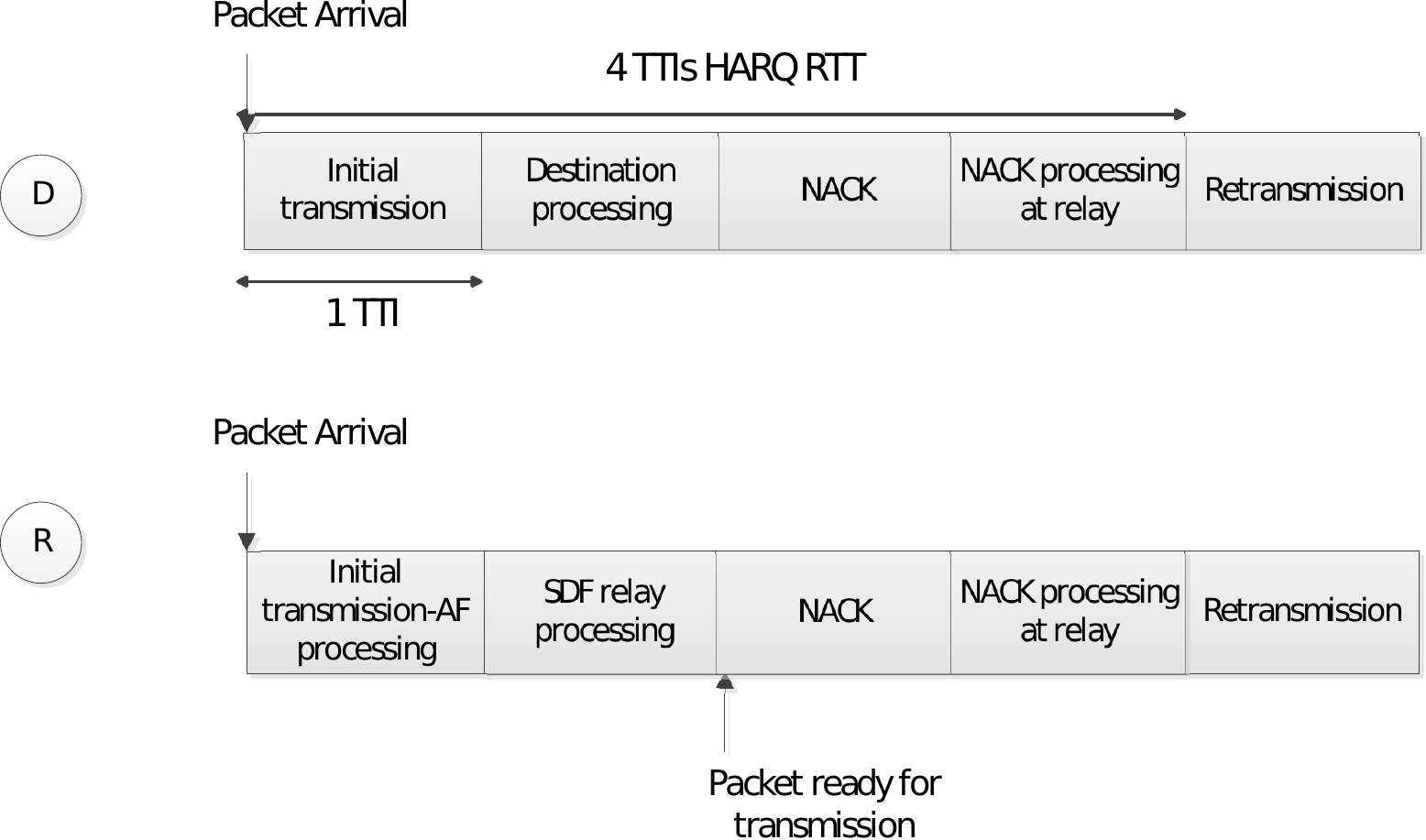}
\par\end{centering}

\protect\caption{RTT diagram of the proposed conventional HARQ procedure.\label{fig:Diagram-of-Relay-based}}
\end{figure}

The additional delay on the transmission due to HARQ is named HARQ
RTT. The HARQ RTT incorporates unavoidable physical delays, such as
processing times, propagation delays, and TTI duration. In \cite{Pacovi2018,strodthoff,Pocovi2017,Pedersen2017,RTT_reduction},
it has been shown that assuming LTE-alike asynchronous HARQ operation
with a minimum RTT of 8 TTIs, even with a very short TTI duration
of 0.143 ms, the HARQ RTT would not satisfy the 1 ms URLLC latency
target. This is the main reason of HARQ abandonment for the 1 ms end-to-end
latency use case of URLLC, at least for the initial URLLC specification
in Rel.15 \cite{rel15}. In this work, in order to allow room for
one HARQ retransmission, the reduced HARQ RTT of 4 TTIs has been adopted
\cite{Pocovi2017}. The proposed HARQ RTT for one FD-relay based system
is presented in Fig. \ref{fig:Diagram-of-Relay-based} %
\footnote{In this work, the propagation delay is neglected. In fact, the propagation
delay depends on the distance between two nodes, however as the latter
is out of the scope of this work, all nodes are supposed to have the
same time offset. %
}. We consider $1$ TTI for the initial transmission, $1$ TTI for
the destination processing required to decode the initial transmission%
\footnote{In parellel with the destination processing, the relay performs the
SDF relay processing of the received packet.%
}, $1$ TTI for the NACK feedback, and $1$ TTI for retransmitter processing
of NACK.

\begin{figure*}[tbh]
\begin{centering}
\includegraphics[scale=0.50]{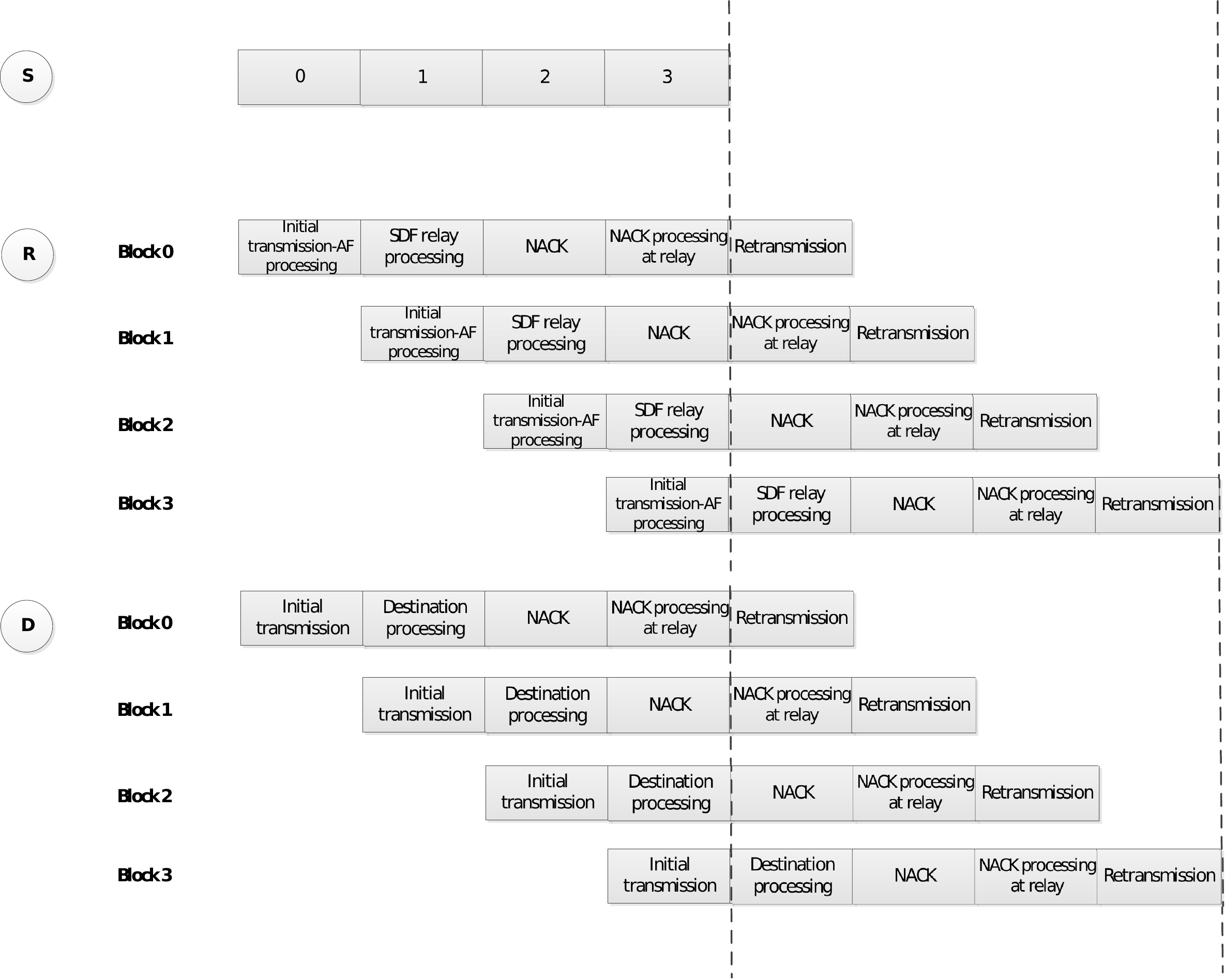}
\par\end{centering}

\protect\caption{RTT diagram of conventional HARQ procedure with multiple HARQ processes.\label{fig:Multiple_Access_HARQ}}
\end{figure*}

In order to allow continuous transmission while previous blocks are
being decoded, four consecutive TTI assignment using multiple HARQ
processes is adopted \cite{rel15SW}. As shown in Fig. \ref{fig:Multiple_Access_HARQ} %
\footnote{The NACK processing and the SDF processing are covered by the baseband processor (BP), while the NACK reception and the data packet transmission/retransmission are performed by the radio frequency (RF) front-end. %
} \footnote{The NACK processing is generally based on repetition code, whereas the SDF processing is based on LDPC or polar codes. Thereby, the NACK processing would be completed earlier than the SDF processing. %
},
the consecutive assignment using multiple HARQ processes allows the
resources to be assigned in consecutive TTIs to the same transmitting
source node. At the relay and destination nodes, simultaneous HARQ
processes operate in parallel to decode consecutive assignments. At
the destination%
\footnote{The ACK/NACKs of received data packets are transmitted on a separate reverse link control channel, i.e., the NR physical uplink control channel (PUCCH), with orthogonal resources to the one in which forward link data packets are transmitted.}, each HARQ process is responsible for decoding one
assignment, and transmitting the associated ACK or NACK $2$ TTIs after
the end of that assignment. By using four consecutive TTI assignment,
the system can achieve a maximum data rate that is four times greater
than the non consecutive assignment if every block is decoded correctly.

\subsubsection{Reliability Issue}

In order to achieve higher reliability, at least one retransmission
is needed over a time varying channel. However, in the conventional
system, presented above, the retransmission is activated only if the
relay can correctly decode the received signal. In the next subsection,
we propose an enhanced HARQ procedure that can solve this reliability
issue.

\subsection{Relay-based Enhanced HARQ Procedure}

\begin{figure}[tbh]
\begin{centering}
\includegraphics[scale=0.50]{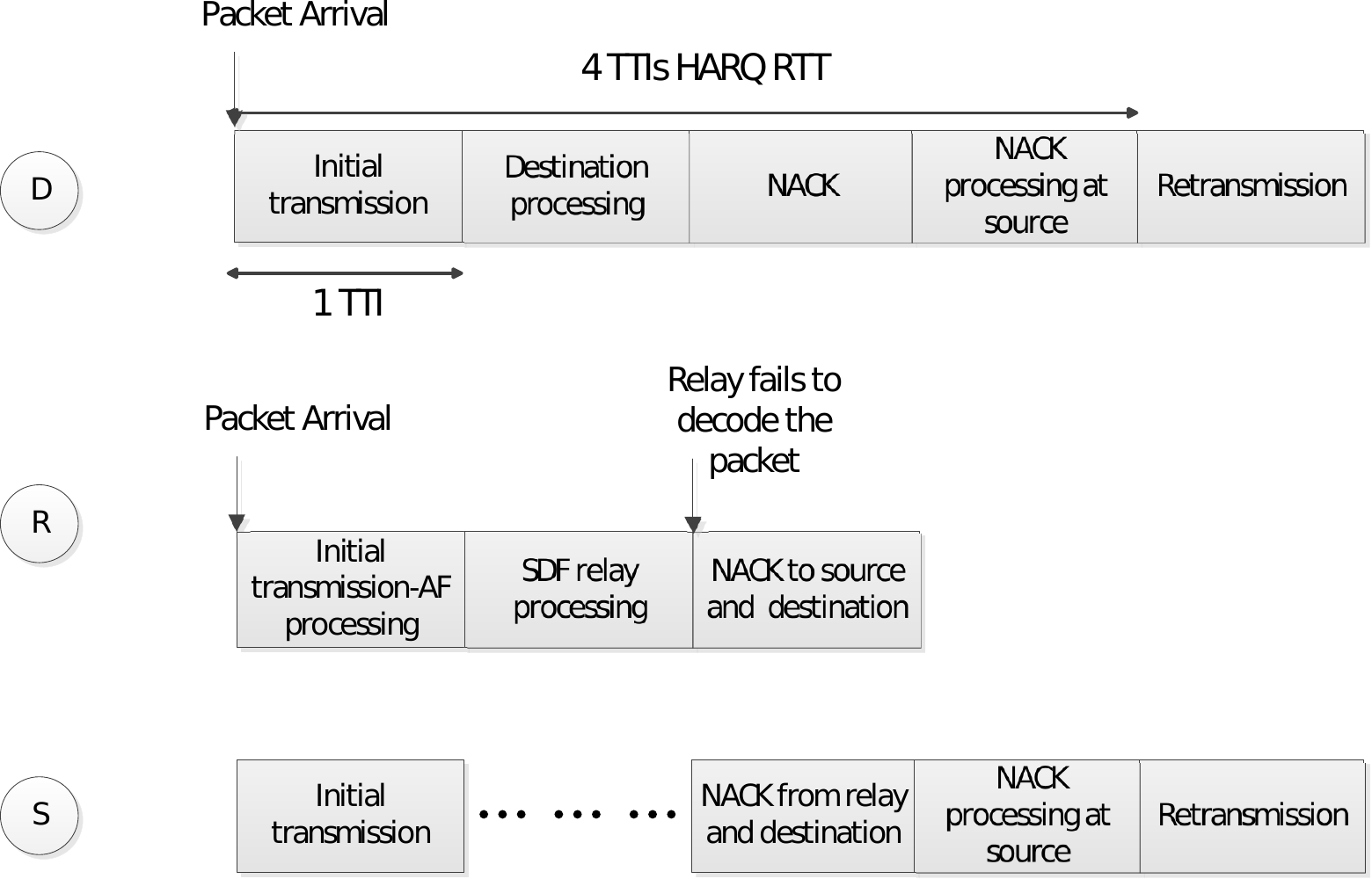}
\par\end{centering}

\protect\caption{RTT diagram of enhanced HARQ procedure.\label{fig:RTT-diagram-Enhanced}}
\end{figure}

In the conventional system, presented above, the retransmission is
activated only if the relay can correctly decode the received signal
which reduces the relay-based HARQ system performance. To solve this
problem, we propose an enhanced  HARQ procedure that makes use of the modified SDF extensively studied in the literature\cite{turbo, Maham, Maaz}. In the proposed procedure,
when the relay decoding outcome is erroneous, the relay broadcasts
a NACK to both the destination and the source to indicate that the
source will play the retransmitter role during Phase III, as shown
in Fig. \ref{fig:RTT-diagram-Enhanced}. In this enhanced procedure,
the received signal at the destination, during Phase III, is expressed
as
\begin{equation}
y_{\mathrm{D}}^{III}(t)=\left(\underset{\mathrm{Retransmitted\, signal}}{\underbrace{\sqrt{P_{\mathrm{e}}}h_{\mathrm{e}}^{III}x_{\mathrm{s}}(t)}}+n_{\mathrm{D}}^{III}(t)\right)\times a,\label{eq:combinDest-2-1-1}
\end{equation}
 with 
\[
a=\begin{cases}
1 & \textrm{upon a reception of NACK from the destination}\\
0 & \textrm{Otherwise},
\end{cases}
\]
 
\[
P_{\mathrm{e}}=\begin{cases}
P_{\textrm{S}} & \textrm{upon a reception of NACK from the relay}\\
P_{\mathrm{R}} & \textrm{Otherwise},
\end{cases}
\]
and 
\[
h_{\mathrm{e}}^{III}=\begin{cases}
h_{\mathrm{SD}}^{III} & \textrm{upon a reception of NACK from the relay}\\
h_{\mathrm{RD}}^{III} & \textrm{Otherwise}.
\end{cases}
\]

\section{\label{sec:Outage-Probability}Outage Probability}

In this section, we derive the outage probability of the proposed
system where packet retransmission is activated only if the destination
fails to decode the transmitted data packet. Note that, the system
outage occurs when both transmissions, i.e., Phase I and Phase III
are in outage. Therefore, the system outage probability is given by

\begin{equation}
\begin{aligned}P_{\mathrm{out}} & =P_{\mathrm{out}}^{I}\times P_{\mathrm{out}}^{III},\end{aligned}
\label{eq:-1}
\end{equation}
where $P_{\mathrm{out}}^{I}$ and $P_{\mathrm{out}}^{III}$ represent
respectively, the system outage probability related to Phase I and
Phase III. Overall, to derive the system outage probability, i.e.,
$P_{\mathrm{out}}$, the received instantaneous SINR of $\mathrm{S}\rightarrow\mathrm{D}$,
$\mathrm{S}\rightarrow\mathrm{R}$, and $\mathrm{R}\rightarrow\mathrm{D}$
links need to be provided. They are, respectively, given as, $\gamma_{\mathrm{SD}}=\frac{P_{\textrm{S}}|h_{\mathrm{SD}}|^{2}}{N_{D}}$,
$\gamma_{\mathrm{RD}}=\frac{P_{\mathrm{R}}|h_{\mathrm{RD}}|^{2}}{N_{D}}$,{\footnotesize{}
} and $\gamma_{\mathrm{SR}}=\frac{P_{\textrm{S}}|h_{\mathrm{\mathrm{SR}}}|^{2}}{P_{\mathrm{R}}\sigma_{\mathrm{RR}}^{2}+N_{R}}$.
Note that all SINRs are exponentially distributed random variables. \\
Indeed, URLLC is known to cope with the finite-blocklength regime. However, for the sake of simplicity, we proceed in this study, with the case of sufficiently long codes since it has been proven to provide very close performance to  the case with finite block-length \cite{zorzi, alouini}.

First, let's derive the outage probability expression, relative to
Phase I as \cite{airod}

\begin{align}
\begin{aligned}P_{out}^{I} & =\mathrm{Pr}\left(\frac{T}{T+\tau}I(\mathbf{x}_{\mathrm{s}_{f}},\mathbf{y_{\mathrm{D}_{f}}^{I}})<R\right)\\
 & =\mathrm{Pr}\left(\frac{1}{T+\tau}{\displaystyle \sum_{i=0}^{T-1}}\log_{2}\left(1+\gamma_{i}^{I}\right)<R\right),
\end{aligned}
\label{eq:OutPh1}
\end{align}
with $R$ is the bit rate per channel use, the factor $\frac{T}{T+\tau}$
means that the transmission of $T$ useful codewords occupies $T+\tau$
channel uses, and $I$ represents the overall AF system average mutual
information which is manipulated as below

{\footnotesize{}
\begin{equation}
\begin{alignedat}{1}I(\mathbf{x}_{\mathrm{s}_{f}},\,\mathbf{y_{\mathrm{D}_{f}}^{I}}) & =\frac{1}{T}\sum_{i=0}^{T-1}\mathrm{log_{2}}(1+\gamma_{i}^{I})\\
 & =\frac{1}{T}\sum_{i=0}^{T-1}\mathrm{log_{2}}\left[(1+\rho_{I})\left(1+\frac{2\left|\mu_{I}\right|cos\left(2\pi i\frac{\tau}{T}+\theta\right)}{1+\rho_{I}}\right)\right]\\
 & =\mathrm{log_{2}}(1+\rho_{I})+\frac{1}{T}\sum_{i=0}^{T-1}\mathrm{log_{2}}\Biggl(1+\Biggl(\frac{2\left|\mu_{I}\right|}{1+\rho_{I}}\\
 & \times\mathrm{cos}\left(2\pi i\frac{\tau}{T}+\theta\right)\Biggr)\Biggr).
\end{alignedat}
\label{eq:Mutual_Info}
\end{equation}
}
According to the arithmetic-geometric mean inequality, $\rho_{I}\geq2\left|\mu_{I}\right|$,
we have {\footnotesize{$1+\rho_{I}>2\left|\mu_{I}\right|cos\left(2\pi i\frac{\tau}{T}+\theta\right)$}}.
Thus, by using the first order Taylor expansion, we have $\mathrm{ln}\left(1+\frac{2\left|\mu_{I}\right|cos\left(2\pi i\frac{\tau}{T}+\theta\right)}{1+\rho_{I}}\right)\thickapprox\frac{2\left|\mu_{I}\right|cos\left(2\pi i\frac{\tau}{T}+\theta\right)}{1+\rho_{I}}.$
Noting that {\small{${\displaystyle \sum_{i=0}^{T-1}}cos\left(2\pi i\frac{\tau}{T}+\theta\right)=0$}},
the mutual information, in (\ref{eq:Mutual_Info}), can be approximated
as

\begin{align}
I(\mathbf{x}_{\mathrm{s}_{f}},\mathbf{y_{\mathrm{D}_{f}}^{I}}) & \thickapprox\mathrm{log_{2}}(1+\rho_{I})
\end{align}
Therefore, (\ref{eq:OutPh1}) can be represented as

\begin{eqnarray}
P_{out}^{I} & \approx & \mathrm{Pr}\left(\frac{T}{T+\tau}\log_{2}\left(1+\rho_{I}\right)<R\right),\nonumber \\
 & = & F_{\rho_{I}}\left(\eta_{I}\right),
\end{eqnarray}
with $F_{\rho_{I}}\left(.\right)$ is the CDF of $\rho_{I}$ and $\eta_{I}=2^{\left(\frac{T+\tau}{T}\right)R}-1$.

By substituting $\mathrm{S}\rightarrow\mathrm{D}$, $\mathrm{S}\rightarrow\mathrm{R}$, 
and $\mathrm{R}\rightarrow\mathrm{D}$ links received instantaneous
SINR into $\rho_{I}$, we get

\begin{equation}
\begin{aligned}\rho_{I}= & \frac{\gamma_{\mathrm{SR}}\gamma_{\mathrm{RD}}+\gamma_{\mathrm{SD}}\gamma_{\mathrm{SR}}+\gamma_{\mathrm{SD}}}{1+\gamma_{\mathrm{SR}}+\gamma_{\mathrm{RD}}}.\end{aligned}
\end{equation}
Thus, the CDF of $\rho_{I}$ can be derived as

\begin{align}
\begin{aligned}F_{\rho_{I}}\left(x\right) & =\mathrm{Pr}(\rho_{I}<x)\\
 & =\mathrm{Pr}\left(\gamma_{\mathrm{SD}}\leq\underset{\mathrm{\Delta}}{\underbrace{\frac{(1+\gamma_{\mathrm{SR}}+\gamma_{\mathrm{RD}})x-\gamma_{\mathrm{SR}}\gamma_{\mathrm{RD}}}{1+\gamma_{\mathrm{SR}}}}}\right)\\
 & =\iint_{\Delta\geq0}F_{\gamma_{\mathrm{SD}}}\left(\Delta\right)\times f_{\gamma_{\mathrm{SR}}}(y)\times f_{\gamma_{\mathrm{RD}}}(z)dydz
\end{aligned}
\label{eq:9}
\end{align}
where $F_{\gamma_{\mathrm{SD}}}\left(\Delta\right)=\mathrm{Pr}(\gamma_{\mathrm{SD}}<\Delta)=1-\textrm{e}^{-\alpha_{\mathrm{SD}}\triangle}$,
$f_{\gamma_{\mathrm{SR}}}(y)=\textrm{\ensuremath{\alpha_{\mathrm{SR}}}e}^{-\alpha_{\mathrm{SR}}y}$, and $f_{\gamma_{\mathrm{RD}}}(z)=\textrm{\ensuremath{\alpha_{\mathrm{RD}}}e}^{-\alpha_{\mathrm{RD}}z}$, with $\alpha_{\mathrm{SD}}=\frac{1}{P_{\mathrm{S}}\sigma_{\mathrm{SD}}^{2}}$,
$\alpha_{\mathrm{SR}}=\frac{1}{P_{\mathrm{S}}\sigma_{\mathrm{SR}}^{2}}$,
and $\alpha_{\mathrm{RD}}=\frac{1}{P_{\mathrm{R}}\sigma_{\mathrm{RD}}^{2}}$.
Then, after some manipulations, (\ref{eq:9}) can be derived as

\begin{equation}
\begin{alignedat}{1}F_{\rho_{I}}\left(\eta_{I}\right) & =1-\alpha_{\mathrm{SR}}\intop_{\eta_{I}}^{\infty}e^{-\alpha_{\mathrm{SR}}y}\times e^{-\alpha_{\mathrm{RD}}\left(\frac{1+y}{y-\eta_{I}}\right)\eta_{I}}dy\\
 & -\alpha_{\mathrm{RD}}\times\alpha_{\mathrm{SR}}\times e^{(-\alpha_{\mathrm{SD}}\eta_{I})}\times\\
 & \Biggl(\intop_{0}^{\infty}\frac{e^{-\alpha_{\mathrm{SR}}y}}{\left(\alpha_{\mathrm{RD}}+\left(\frac{\eta_{I}-y}{1+y}\right)\alpha_{\mathrm{SD}}\right)}dy-\\
 & e^{-\alpha_{\mathrm{SD}}\eta_{I}}\intop_{\eta_{I}}^{\infty}\frac{e^{-\left(\alpha_{\mathrm{SR}}y+\alpha_{\mathrm{RD}}\left(\frac{1+y}{y-\eta_{I}}\right)\eta_{I}\right)}}{\left(\alpha_{\mathrm{RD}}+\left(\frac{\eta_{I}-y}{1+y}\right)\alpha_{\mathrm{SD}}\right)}dy\Biggr).
\end{alignedat}
\label{eq:out_phase1}
\end{equation}
Note that all integrals in (\ref{eq:out_phase1}) do not generate
a closed form expression, but can be evaluated numerically using Matlab
software.

The Phase III outage probability $P_{\mathrm{out}}^{III}$ depends
on the adopted HARQ procedure. In general, the $P_{\mathrm{out}}^{III}$
can be expressed as \cite{wcnc18}

\begin{equation}
\begin{aligned}P_{\mathrm{out}}^{III} & =P_{\mathrm{out}}^{\mathrm{S\rightarrow R}}\mathcal{X}+P_{\mathrm{out}}^{\mathrm{S,R\rightarrow D}}\left(1-P_{\mathrm{out}}^{\mathrm{S\rightarrow R}}\right),\end{aligned}
\label{eq:out_syst}
\end{equation}
where $P_{out}^{\mathrm{S\rightarrow R}}$ denotes, the outage probability
of $\mathrm{S\rightarrow R}$ link, and can be expressed as
\begin{eqnarray}
P_{out}^{\mathrm{S\rightarrow R}}= & \mathrm{Pr}(\gamma_{\mathrm{SR}}<\eta)= & 1-\textrm{e}^{-\frac{\eta\left(P_{\mathrm{R}}\sigma_{\mathrm{RR}}^{2}+1\right)}{P_{\textrm{S}}\sigma_{\mathrm{SR}}^{2}}},\label{eq:SR_SD-1}
\end{eqnarray}
 where $\eta=2^{R}-1$. In the conventional HARQ procedure, the only
retransmitter node is the relay and the retransmission happens only
if the $\mathrm{S}\rightarrow\mathrm{R}$ link is not in outage. Therefore,
$\mathcal{X}=1$ and $P_{\mathrm{out}}^{\mathrm{S,R\rightarrow D}}$
represents the outage probability of the combined signal, i.e., Phase
I and Phase III received signals, at the destination side when the
relay is the retransmitter node. In the enhanced  HARQ procedure,
when the $\mathrm{S}\rightarrow\mathrm{R}$ link is in outage, the
source plays the role of the retransmitter instead of the relay during
Phase III. Consequently, $\mathcal{X}=P_{\mathrm{out}}^{\mathrm{S,S\rightarrow D}}$
with $P_{\mathrm{out}}^{\mathrm{S,S\rightarrow D}}$ represents the
outage probability of the combined signal, at the destination side
when the source is the retransmitter node.

In multiple HARQ processes system, when one block $u$ ($0\leq u\leq3$)
is correctly decoded at the destination side, its retransmission TTI
$u+4$ is used for the initial transmission of the next block $v$
in the buffer ($4\leq v\leq7$). However, for the sake of simplicity,
we assume that the TTI $u+4$ is allocated for packet retransmission
even if Phase III is deactivated. Therefore, the $8$-TTI relaying
system can be viewed as a repetition coding scheme where $8$ parallel
sub-channels are used to transmit $4$ symbols message \cite[p.194]{Tse}.
The outage probability $P_{\mathrm{out}}^{\mathrm{S,R\rightarrow D}}$
in (\ref{eq:out_syst}) can be derived as follows
\begin{align}
P_{\mathrm{out}}^{\mathrm{S,R\rightarrow D}} & =\mathrm{Pr}\left(\frac{1}{2T}{\displaystyle \sum_{i=0}^{T-1}}\log_{2}\left(1+\gamma_{i}^{III}\right)<R\right),\nonumber \\
 & \approx\mathrm{Pr}\left(\frac{1}{2}\log_{2}\left(1+\rho_{III}\right)<R\right),\label{eq:SRD outage}
\end{align}
 with $\rho_{III}$ is given by

\begin{equation}
\rho_{III}=\frac{\gamma_{\mathrm{SR}}\gamma_{\mathrm{RD}}+\gamma_{\mathrm{SD}}\gamma_{\mathrm{SR}}+\gamma_{\mathrm{SD}}}{1+\gamma_{\mathrm{SR}}+\gamma_{\mathrm{RD}}}+\gamma_{\mathrm{RD}},
\end{equation}
thereby, $P_{\mathrm{out}}^{\mathrm{S,R\rightarrow D}}$ can be derived
as

{\footnotesize{}
\begin{align}
\begin{aligned}\mathrm{Pr}(\rho_{III}<x) & =\mathrm{Pr}\left(\gamma_{\mathrm{SD}}\leq\underset{\mathrm{\varphi}}{\underbrace{\frac{(1+\gamma_{\mathrm{SR}}+\gamma_{\mathrm{RD}})(x-\gamma_{\mathrm{RD}})-\gamma_{\mathrm{SR}}\gamma_{\mathrm{RD}}}{1+\gamma_{\mathrm{SR}}}}}\right)\\
 & =\iint_{\varphi\geq0}F_{\mathrm{SD}}\left(\varphi\right)\times f_{\gamma_{\mathrm{SR}}}(y)\times f_{\gamma_{\mathrm{RD}}}(z)dydz.
\end{aligned}
\label{eq:19-1-1}
\end{align}
}{\footnotesize \par}

To proceed, let's denote $g(\gamma_{\mathrm{SR}},z)=\left(1-\textrm{e}^{-\alpha_{\mathrm{SD}}\varphi}\right)\times f_{\gamma_{\mathrm{RD}}}(z)$.
First, we have to integrate $g(\gamma_{\mathrm{SR}},z)$ while considering
$\gamma_{\mathrm{SR}}$ as constant. Thereby, we extract limit bounds
for z variable in a way to satisfy $\left((1+\gamma_{\mathrm{SR}}+\gamma_{\mathrm{RD}})(x-\gamma_{\mathrm{RD}})-\gamma_{\mathrm{SR}}\gamma_{\mathrm{RD}}\right)>0$,
after some manipulations we found so, the corresponding bounds, i.e.,
$z\in\left[0,\,\frac{\sqrt{\left(\left(2\gamma_{\mathrm{SR}}+1-x\right)^{2}+4x(\gamma_{\mathrm{SR}}+1)\right)}+\left(x-\left(2\gamma_{\mathrm{SR}}+1\right)\right)}{2}\right],\,\forall\gamma_{\mathrm{SR}}$.
Applying \cite[2.33.3*]{tabl_int52}, we get therefore, the first
integral resolution, then, we integrate the resulting expression,
depending to the variable $\gamma_{\mathrm{SR}}$, i.e., $y\in\left[0,\,\infty\right]$,
to get thus, the system outage probability as in (\ref{eq:out_phase3}),
with $\eta_{III}=2^{2R}-1$ and $erfi$ represents the imaginary error function. 

\begin{flushleft}
{\small{}}
\begin{algorithm*}[tbh]

\begin{raggedright}
{\small{}
\begin{equation}
\begin{aligned}P_{\mathrm{out}}^{\mathrm{S,R\rightarrow D}} & =1-\alpha_{\mathrm{SR}}e^{-\frac{\left(\eta_{III}-1\right)\alpha_{\mathrm{RD}}}{2}}\intop_{0}^{\infty}\left(e^{-\left(\left(\alpha_{\mathrm{SR}}-\alpha_{\mathrm{RD}}\right)y+\alpha_{\mathrm{RD}}\left(\frac{\sqrt{\left(\left(2y+1-\eta_{III}\right)^{2}+4\eta_{III}(y+1)\right)}}{2}\right)\right)}\right)\, dy\\ & -\frac{\alpha_{\mathrm{SR}}\alpha_{\mathrm{RD}}e^{-\alpha_{\mathrm{SD}}\eta_{III}}}{2}\intop_{0}^{\infty}\Biggl(\sqrt{\frac{\pi\left(y+1\right)}{\alpha_{\mathrm{SD}}}}\times e^{-\left(\frac{\left(2y+1-\eta_{III}+\alpha_{\mathrm{RD}}\right)^{2}}{y+1}\right)}\times\textrm{e}^{-\alpha_{\mathrm{SR}}y}\\ & \times\Biggl(\mathrm{erfi}\Biggl(\sqrt{\frac{\alpha_{\mathrm{SD}}}{y+1}}\times\Biggl(\left(y-\frac{(\eta_{III}-1)}{2}\right)+\frac{\sqrt{\left(\left(2y+1-\eta_{III}+\alpha_{\mathrm{RD}}\right)^{2}+4\eta_{III}(y+1)\right)}}{2}\Biggr)\Biggr)\\ & -\mathrm{erfi}\Biggl(\sqrt{\frac{\alpha_{\mathrm{SD}}}{y+1}}\times\mathrm{\left(2y+1-\mathit{\eta_{III}+\alpha_{\mathrm{RD}}}\right)\Biggr)}\Biggr)\Biggr)\, dy,\end{aligned}
\label{eq:out_phase3}
\end{equation}
}
\par\end{raggedright}{\small \par}

\end{algorithm*}

\par\end{flushleft}{\small \par}

\begin{flushleft}
The outage probability $P_{\mathrm{out}}^{\mathrm{S,S\rightarrow D}}$
in (\ref{eq:out_syst}) is given by
\begin{alignat}{1}
P_{\mathrm{out}}^{\mathrm{S,S\rightarrow D}} & \approx\mathrm{Pr}\left(\frac{\gamma_{\mathrm{SR}}\gamma_{\mathrm{RD}}+\gamma_{\mathrm{SD}}\gamma_{\mathrm{SR}}+\gamma_{\mathrm{SD}}}{1+\gamma_{\mathrm{SR}}+\gamma_{\mathrm{RD}}}+\gamma_{\mathrm{SD}}<\eta_{III}\right),\label{eq:outage-SSD}
\end{alignat}
we get then
\par\end{flushleft}

{\small{}
\begin{equation}
\begin{aligned}P_{\mathrm{out}}^{\mathrm{S,S\rightarrow D}} & =\textrm{Pr}\left(\underset{\mathrm{\rho_{I}}}{\underbrace{\frac{\gamma_{\mathrm{SR}}\gamma_{\mathrm{RD}}+\gamma_{\mathrm{SD}}\gamma_{\mathrm{SR}}+\gamma_{\mathrm{SD}}}{1+\gamma_{\mathrm{SR}}+\gamma_{\mathrm{RD}}}}}\leq\underset{\mathrm{\Psi}}{\underbrace{\left(\eta_{III}-\gamma_{\mathrm{SD}}\right)}}\right)\\
 & =\int_{\Psi\geq0}F_{\rho_{I}}\left(\Psi\right)\times f_{\gamma_{\mathrm{SD}}}(x)dx.
\end{aligned}
\end{equation}
}Let's first develop the first term in (\ref{eq:out_phase1}), $X=1-\alpha_{\mathrm{SR}}\intop_{\eta_{I}}^{\infty}e^{-\alpha_{\mathrm{SR}}y}\times e^{-\alpha_{\mathrm{RD}}\left(\frac{1+y}{y-\eta_{I}}\right)\eta_{I}}dy$,
using the integration by change of variables, $u=y-\eta_{I}$, while
applying \cite[3.471.9]{tabl_int52}, we get

{\small{}
\begin{align}
\begin{alignedat}{1}X & =1-\alpha_{\mathrm{SR}}e^{-(\alpha_{\mathrm{SR}+}\alpha_{\mathrm{RD}})\eta_{I}}\intop_{0}^{\infty}e^{-\alpha_{\mathrm{RD}}\left(\frac{1+\eta_{I}+u}{u}\right)\eta_{I}-\alpha_{\mathrm{SR}}u}du\\
 & =1-\Biggl(2e^{-(\alpha_{\mathrm{SR}+}\alpha_{\mathrm{RD}})\eta_{I}}\sqrt{\alpha_{\mathrm{SR}}\alpha_{\mathrm{RD}}\left(1+\eta_{I}\right)\eta_{I}}\\
 & \times K_{1}\left(2\sqrt{\alpha_{\mathrm{SR}}\alpha_{\mathrm{RD}}\left(1+\eta_{I}\right)\eta_{I}}\right)\Biggr)
\end{alignedat}
\label{eq:firstTerm}
\end{align}
}Next, we insert (\ref{eq:firstTerm}) into (\ref{eq:out_phase1}).
Then, we integrate $\left[F_{\rho_{I}}\left(\eta_{III}-x\right)\times f_{\gamma_{\mathrm{SD}}}(x)\right]$
with respect to the variable $\gamma_{\mathrm{SD}}$, i.e., $x$,
that shall fall within the interval defined by $\left[0,\,\eta_{III}\right]$.
We get thus, the system outage probability as in (\ref{eq:System_CDF_sdPhase3}). With $A(x)=\alpha_{\mathrm{SR}}\alpha_{\mathrm{RD}}\left(1+\eta_{III}-x\right)\left(\eta_{III}-x\right).$

\begin{algorithm*}[tbh]
{\small{}
\begin{align}
\begin{aligned}P_{U}^{\mathrm{S,S\rightarrow D}} & =1-e^{-\alpha_{\mathrm{SD}}\eta_{III}}-2\alpha_{\mathrm{SD}}e^{-(\alpha_{\mathrm{SR}+}\alpha_{\mathrm{RD}})\eta_{III}}\intop_{0}^{\eta_{III}}\sqrt{A(x)}\times K_{1}\left(2\sqrt{A(x)}\right)\times e^{-\left(\alpha_{\mathrm{SD}}-\left(\alpha_{\mathrm{SR}+}\alpha_{\mathrm{RD}}\right)\right)x}\, dx\\
 & -\alpha_{\mathrm{SR}}\alpha_{\mathrm{RD}}e^{-\alpha_{\mathrm{SD}}\eta_{III}}\Biggl(\intop_{0}^{\eta_{III}}\Biggl(\intop_{0}^{\infty}\frac{e^{-\left(\alpha_{\mathrm{SR}}y-\alpha_{\mathrm{SD}}x\right)}}{\left(\alpha_{\mathrm{RD}}+\left(\frac{\eta_{III}-y-x}{1+y}\right)\alpha_{\mathrm{SD}}\right)}dy-\\
 & e^{-\alpha_{\mathrm{SD}}\eta_{III}}\intop_{\eta_{III}-x}^{\infty}\frac{e^{-\left(\alpha_{\mathrm{SR}}y+\alpha_{\mathrm{RD}}\left(\frac{1+y}{y-\eta_{III}+x}\right)\left(\eta_{III}-x\right)+2\alpha_{\mathrm{SD}}x\right)}}{\left(\alpha_{\mathrm{RD}}+\left(\frac{\eta_{III}-y-x}{1+y}\right)\alpha_{\mathrm{SD}}\right)}dy\Biggr)dx\Biggr)
\end{aligned}
\label{eq:System_CDF_sdPhase3}
\end{align}
}{\small \par}\end{algorithm*}
{\small \par}Integrals in (\ref{eq:System_CDF_sdPhase3}) do not generate a closed
form expression, but can be evaluated numerically using Matlab software.
Finally, by substituting (\ref{eq:out_phase1}), (\ref{eq:SR_SD-1}),
(\ref{eq:out_phase3}), and (\ref{eq:System_CDF_sdPhase3}) into (\ref{eq:out_syst}),
we get the expression of the system outage probability.

\section{\label{sec:Numerical-Results}Numerical Results}

In this section, the performance of the proposed relay-based HARQ procedures are studied through numerical results, and are labeled respectively as, "conventional sim" and "enhanced sim" in all figures. We used for that purpose, Monte-Carlo simulations to confirm theoretical findings derived in Section
\ref{sec:Outage-Probability}. We then represent the performance simulations
in comparison with the two main relaying schemes with no retransmission
procedure, i.e., the AF as the relaying scheme that includes a very
negligibile latency and the SDF with a latency of $3$ times the transmission
time interval, i.e., $3\times TTI$. We also use the S2D HARQ-based
system where no relay assists the communication between the source
and the destination as a reference to evaluate the performance of
the proposed relay-assisted HARQ procedures %
\footnote{Similar to the proposed system, in the S2D HARQ-based system, only
one HARQ retransmission is allowed.%
}. We assumed for simulation purposes, an equal power allocation among
the source and the relay, i.e., $P=P_{\mathrm{s}}=P_{\mathrm{R}}$,
and $T_{TTI}=125\,\textrm{\ensuremath{\mu}s}$%
\footnote{To achieve a target of 1 ms latency's requirement, we consider a frame
structure for 5G NR that offers a shorten duration of the TTI. Thus
according to a 3rd Generation Partnership Project 3GPP, using the
scalable OFDM numerology, while assuming a subcarrier spacing of 120$\,\textrm{KHZ}$
\cite{zaidi61_latency,70}, the $14$ symbols TTI duration is $T_{\textrm{TTI}}=125\,\textrm{\ensuremath{\mu}s}$
.%
}. The relay and the destination noise powers are assumed to be normalized
to $1$. For a fair comparison, we have set the sum of the transmitted power from the source and the relay nodes of the cooperative system equal to that of
the non-cooperative system (S2D). We note that, $\sigma_{\mathrm{SD}}^{2}$, $\sigma_{\mathrm{SR}}^{2}$, $\sigma_{\mathrm{RD}}^{2}$, and $\sigma_{\mathrm{RR}}^{2}$ stand respectively for, the source-to-destination link gain, the source-to-relay link gain, the relay-to-destination link gain, and the RSI level.

\begin{figure}[h]
\centering{}\includegraphics[scale=0.50]{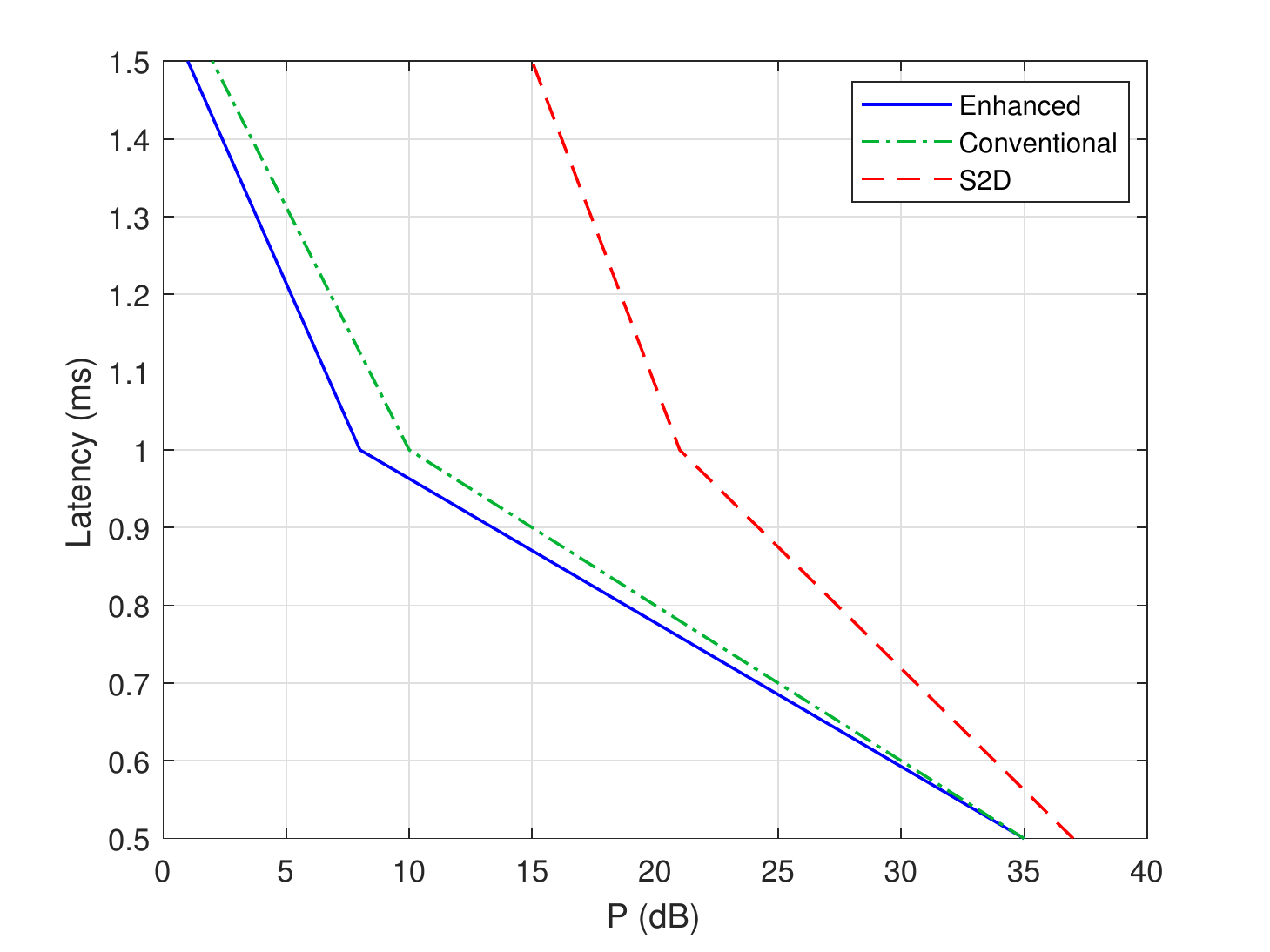}\protect 
\caption{\label{fig:Lat-versusSNRdB} \small \textbf{The latency budget at satisfied level of reliability.}
\\$\sigma_{\mathrm{SR}}^{2}=\sigma_{\mathrm{RD}}^{2}=10\,\mathrm{dB}$,
$\sigma_{\mathrm{SD}}^{2}=0\,\mathrm{dB}$, and negligible RSI. }

\end{figure}
\begin{figure}[H]
\begin{centering}
\includegraphics[scale=0.50]{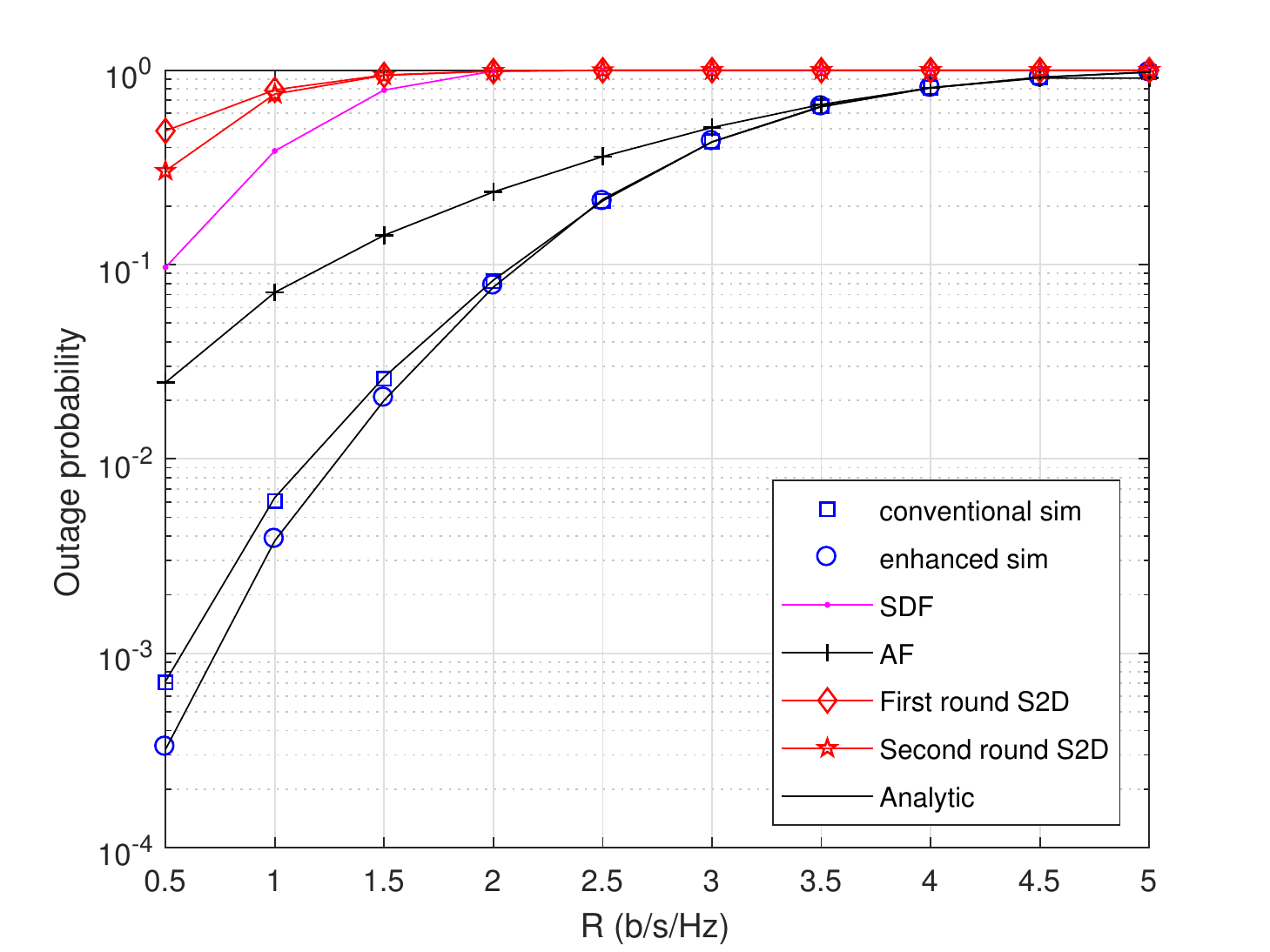}
\par\end{centering}

$\hphantom{\;}\hphantom{\;}\hphantom{\;}\hphantom{\;}\hphantom{\;}\hphantom{\;}\hphantom{\;}\hphantom{\;}\hphantom{\;}\hphantom{\;}\hphantom{\;}\hphantom{\;}\hphantom{\;}\hphantom{\;}\hphantom{\;}\hphantom{\;}\hphantom{\;}\hphantom{\;}\hphantom{\;}\hphantom{\;}\hphantom{\;}\hphantom{\;}\hphantom{\;} a)${\small{}
Outage probability versus} $R$.

\begin{centering}
\includegraphics[scale=0.50]{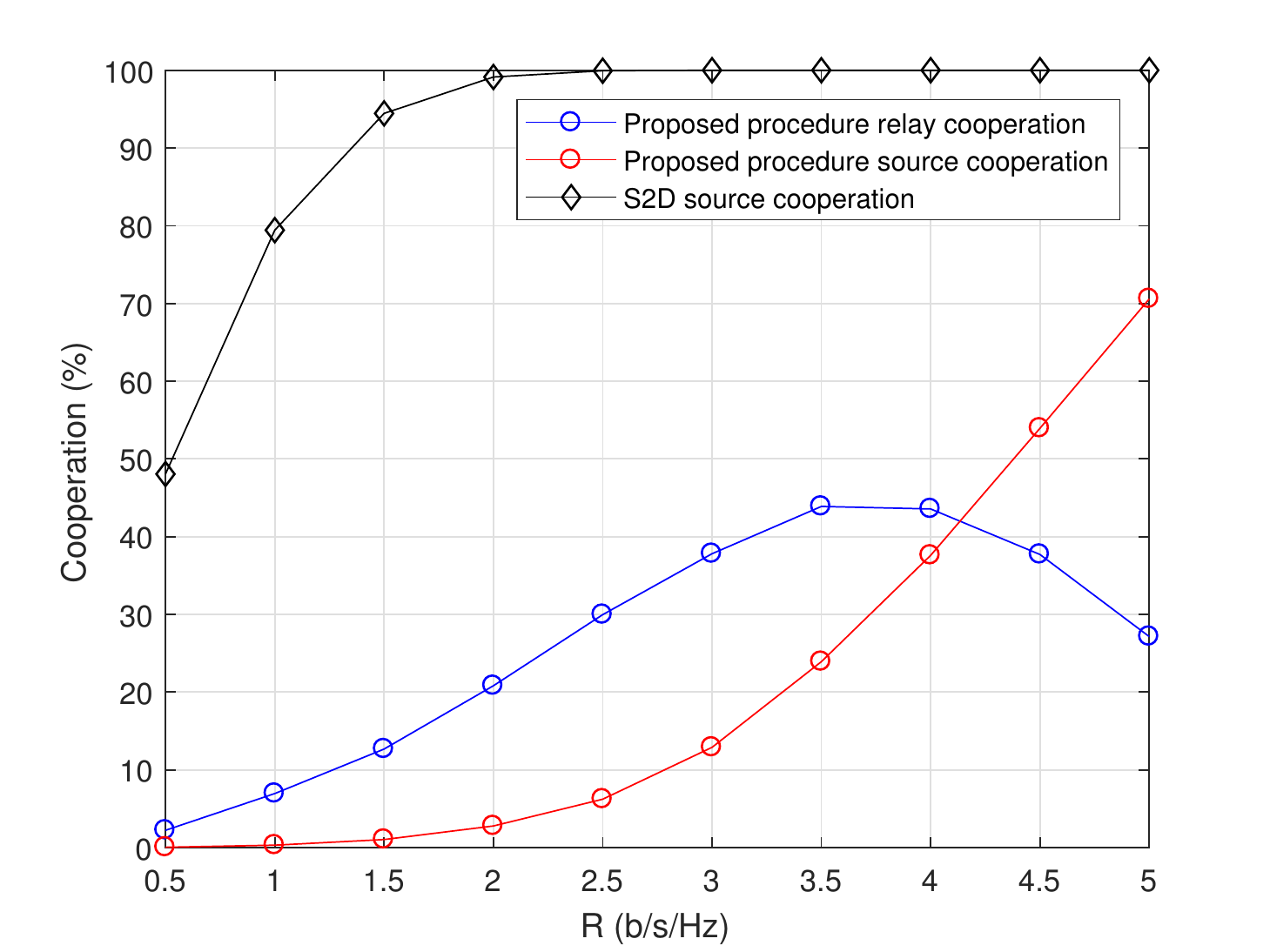} 
\par\end{centering}

$\hphantom{\;}\hphantom{\;}\hphantom{\;}\hphantom{\;}\hphantom{\;}\hphantom{\;}\hphantom{\;}\hphantom{\;}\hphantom{\;}\hphantom{\;}\hphantom{\;}\hphantom{\;}\hphantom{\;}\hphantom{\;}\hphantom{\;}\hphantom{\;}\hphantom{\;}\hphantom{\;}\hphantom{\;}\hphantom{\;}\hphantom{\;}\hphantom{\;}\hphantom{\;} b)$ {\small{}Percentage
of retransmission.}{\small \par}

\protect\caption{\label{fig:Outage-probabilityhigh-versusSD}\small \textbf{The impact of transmission rate for negligible direct link}. 
\\  $\sigma_{\mathrm{SR}}^{2}=\sigma_{\mathrm{RD}}^{2}=10\,\mathrm{dB}$,
$\sigma_{\mathrm{SD}}^{2}=-10\,\mathrm{dB}$, $\sigma_{\mathrm{RR}}^{2}=-10\,\mathrm{dB}$,
and $P=5\,\mathrm{dB}$. }

\end{figure}
\begin{figure}[h]
\begin{centering}
\includegraphics[scale=0.50]{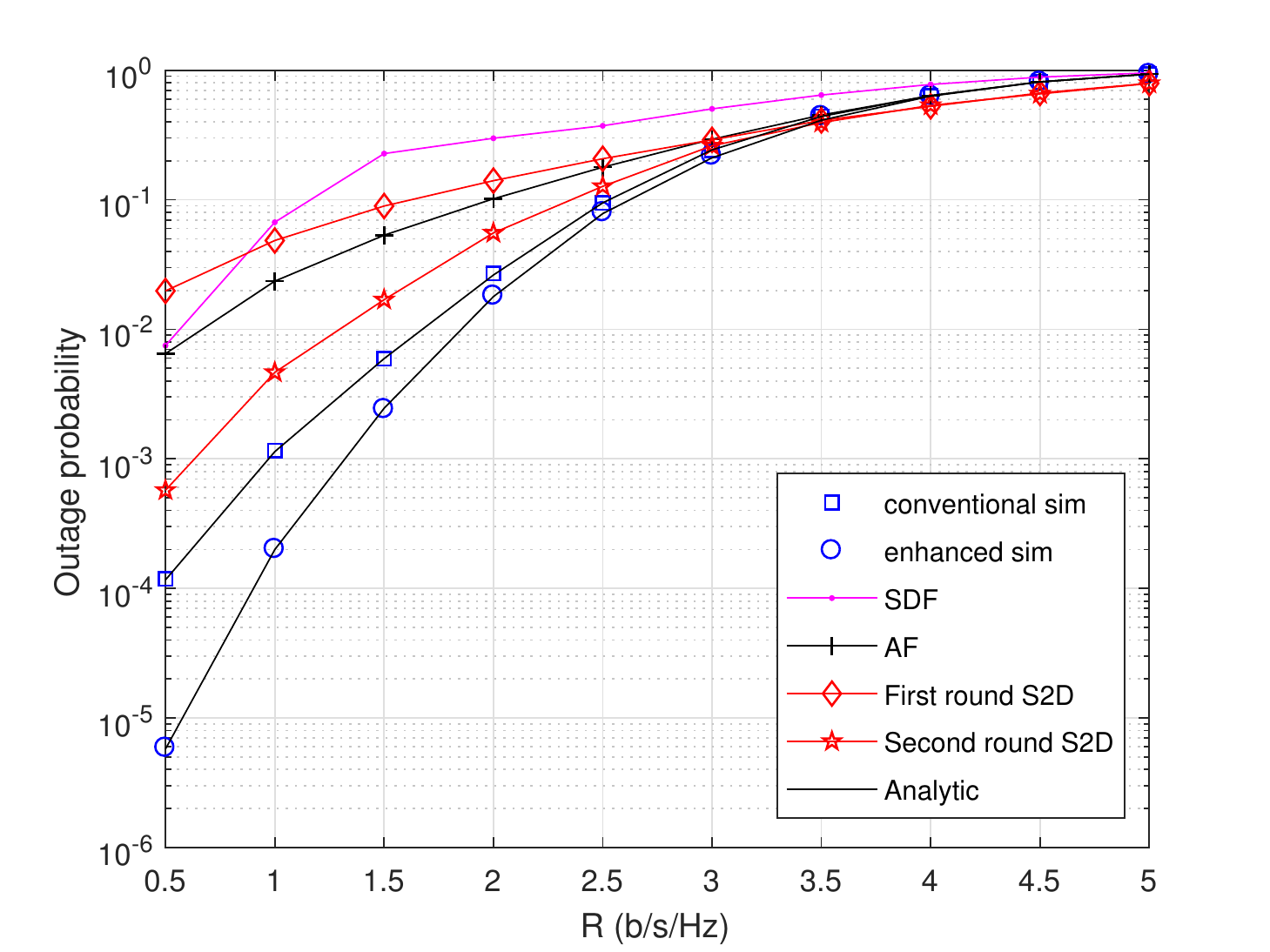}
\par\end{centering}

$\hphantom{\;}\hphantom{\;}\hphantom{\;}\hphantom{\;}\hphantom{\;}\hphantom{\;}\hphantom{\;}\hphantom{\;}\hphantom{\;}\hphantom{\;}\hphantom{\;}\hphantom{\;}\hphantom{\;}\hphantom{\;}\hphantom{\;}\hphantom{\;}\hphantom{\;}\hphantom{\;}\hphantom{\;}\hphantom{\;}\hphantom{\;}\hphantom{\;}\hphantom{\;} a)$
{\small{}Outage probability versus }$R$.

\begin{centering}
\includegraphics[scale=0.50]{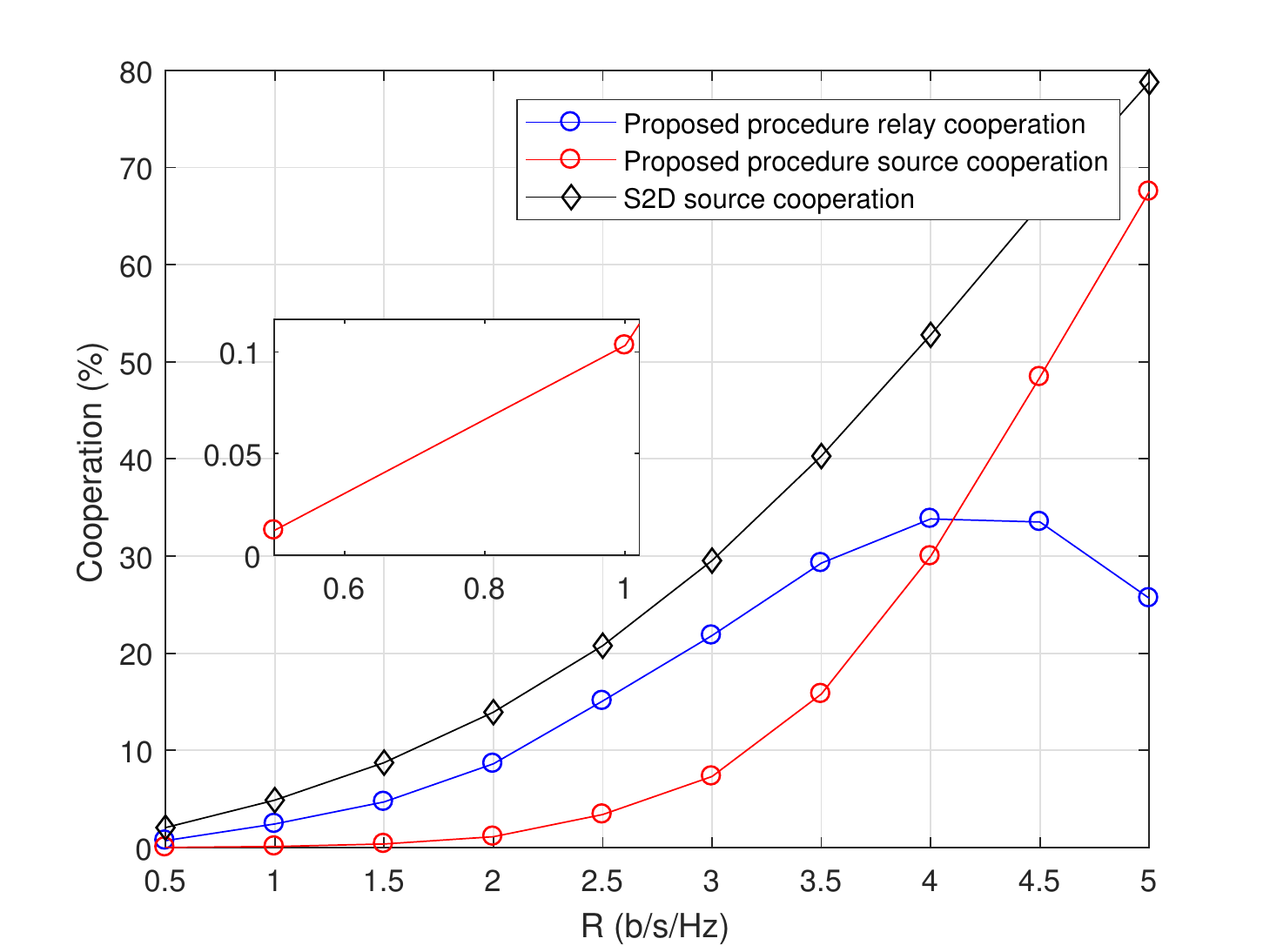}
\par\end{centering}

$\hphantom{\;}\hphantom{\;}\hphantom{\;}\hphantom{\;}\hphantom{\;}\hphantom{\;}\hphantom{\;}\hphantom{\;}\hphantom{\;}\hphantom{\;}\hphantom{\;}\hphantom{\;}\hphantom{\;}\hphantom{\;}\hphantom{\;}\hphantom{\;}\hphantom{\;}\hphantom{\;}\hphantom{\;}\hphantom{\;}\hphantom{\;}\hphantom{\;}\hphantom{\;} b)$ {\small{}Percentage
of retransmission.}{\small \par}

\protect\caption{\label{fig:Outage-probabilitylow-versusSD} \small \textbf{The impact of transmission rate for moderate direct link}. 
\\$\sigma_{\mathrm{SR}}^{2}=\sigma_{\mathrm{RD}}^{2}=10\,\mathrm{dB}$,
$\sigma_{\mathrm{SD}}^{2}=5\,\mathrm{dB}$, $\sigma_{\mathrm{RR}}^{2}=-10\,\mathrm{dB}$,
and $P=5\,\mathrm{dB}$. }
\end{figure}

First, through Fig. \ref{fig:Lat-versusSNRdB}, we assert enhancements noted with both proposed procedures in term of latency. We  represent indeed, the latency level versus the transmit SNR, at a satisfied level of reliability. We compare improvements respectively between, the enhanced procedure, the conventional procedure, and the direct transmission, i.e., S2D. Note that, for the sake of clarity, results are given with a relaxed latency constraint of 1.5 ms, i.e., two allowed retransmissions. The aim is to affirm the flexibility of both procedures as they maintain their gain even with more than one retransmission. As clearly noticed, a significant enhancement is noted with what we propose, specifically, with the enhanced procedure. It is observed that the 1 ms latency constraint with a satisfied level of reliability can be achieved at less than 10 dB for the enhanced procedure whereas it requires more than 20 dB for the S2D scheme. The purpose of this work is to satisfy the 1 ms use case of URLLC, thus,  all outage performance results below, are obtained on the basis of one allowed retansmission.\\
\\
At first, we notice overall, that simulation results match perfectly
with theoretical analysis obtained in Section$\,$\ref{sec:Outage-Probability}. Fig.   
\ref{fig:Outage-probabilityhigh-versusSD}$.a$
and Fig. \ref{fig:Outage-probabilitylow-versusSD}$.a$, depict the
outage performance with respect to data transmission rate, $R$. In
Fig. \ref{fig:Outage-probabilityhigh-versusSD}$.a$, we clearly see 
that when the direct link between the source and the destination is
 negligible ($\sigma_{\mathrm{SD}}^{2}=-10\,\textrm{dB}$), the
S2D transmission fails even after two transmission rounds and using
a relay with better transmission links becomes mandatory. Due to the
SDF relaying high latency, the AF relaying offers the best performance
for one round transmissions. Using a second round transmission in
the relay-based system boosts the cooperative diversity and leads
to a better performance. In Fig. \ref{fig:Outage-probabilitylow-versusSD}$.a$,
we clearly see  that implementing HARQ procedure in a relay based system
enhances the system outage probability from $2\times10^{-2}$ to $2\times10^{-4}$
at $R=1$. Moreover, we notice that while for low direct link gains
($\sigma_{\mathrm{SD}}^{2}=-10\,\textrm{dB}$), the proposed HARQ
procedures have almost similar performance, the enhanced HARQ procedure
tends to outperform the conventional one as the direct link becomes
prominent ($\sigma_{\mathrm{SD}}^{2}=5\,\textrm{dB}$).

In Fig. \ref{fig:Outage-probabilityhigh-versusSD}$.b$ and Fig. \ref{fig:Outage-probabilitylow-versusSD}$.b$,
we plot the percentage of each node cooperation during the second
round when the destination fails to decode the first round transmission.
For S2D system, the percentage of the source cooperation during the
second round is given by $P_{\mathrm{out}}^{\mathrm{S\rightarrow D}}\%$.
For relay-assisted system, the percentage of the relay cooperation
during the second round is given by $\left(1-P_{\mathrm{out}}^{\mathrm{S\rightarrow R}}\right)\times P_{\mathrm{out}}^{I}\%$
while the percentage of the source cooperation during the second round
for enhanced procedure is given by $P_{\mathrm{out}}^{\mathrm{S\rightarrow R}}\times P_{\mathrm{out}}^{I}\%$.

As observed from Fig. \ref{fig:Outage-probabilityhigh-versusSD}$.b$,
in a relay-based system, the source cooperation steadily increases
with the increase of transmission rate starting from zero cooperation
at very low rates. On the other hand, the relay cooperation gives
different performance trends. At low transmission rates, the relay cooperation
increases with the increase of $R$. After some point, it starts decreasing
until it reaches zero cooperation at high rates. This is due to the
fact that, at very low rates the destination can correctly decode
the first round received packet. As the source increases its information
rate, the first round transmission outage takes place with high probability
and thereby the relay is activated to retransmit the packet. As the
source further increases its rate, the $\mathrm{S\rightarrow R}$
link outage occurs more frequently which means that the relay becomes
unable to cooperate using SDF and the system has either to keep silent
during the second round when using the conventional HARQ procedure
or to rely more on the source cooperation in the case of enhanced
HARQ procedure. In Fig. \ref{fig:Outage-probabilitylow-versusSD}$.b$,
when the direct link is non negligible ($\sigma_{\mathrm{SD}}^{2}=5\,\textrm{dB}$),
the S2D source cooperation curve during the second round has same
shape as the relay-based source cooperation curve. However, thanks
to the good relaying links ($\sigma_{\mathrm{SR}}^{2}=\sigma_{\mathrm{RD}}^{2}=10\,\mathrm{dB}$)
the relay-based system outage probability is very low during the first
round compared to the S2D system and thereby the system nodes cooperation
is reduced during the second round with a better outage performance.

\begin{figure}[tbh]
\begin{centering}
\includegraphics[scale=0.50]{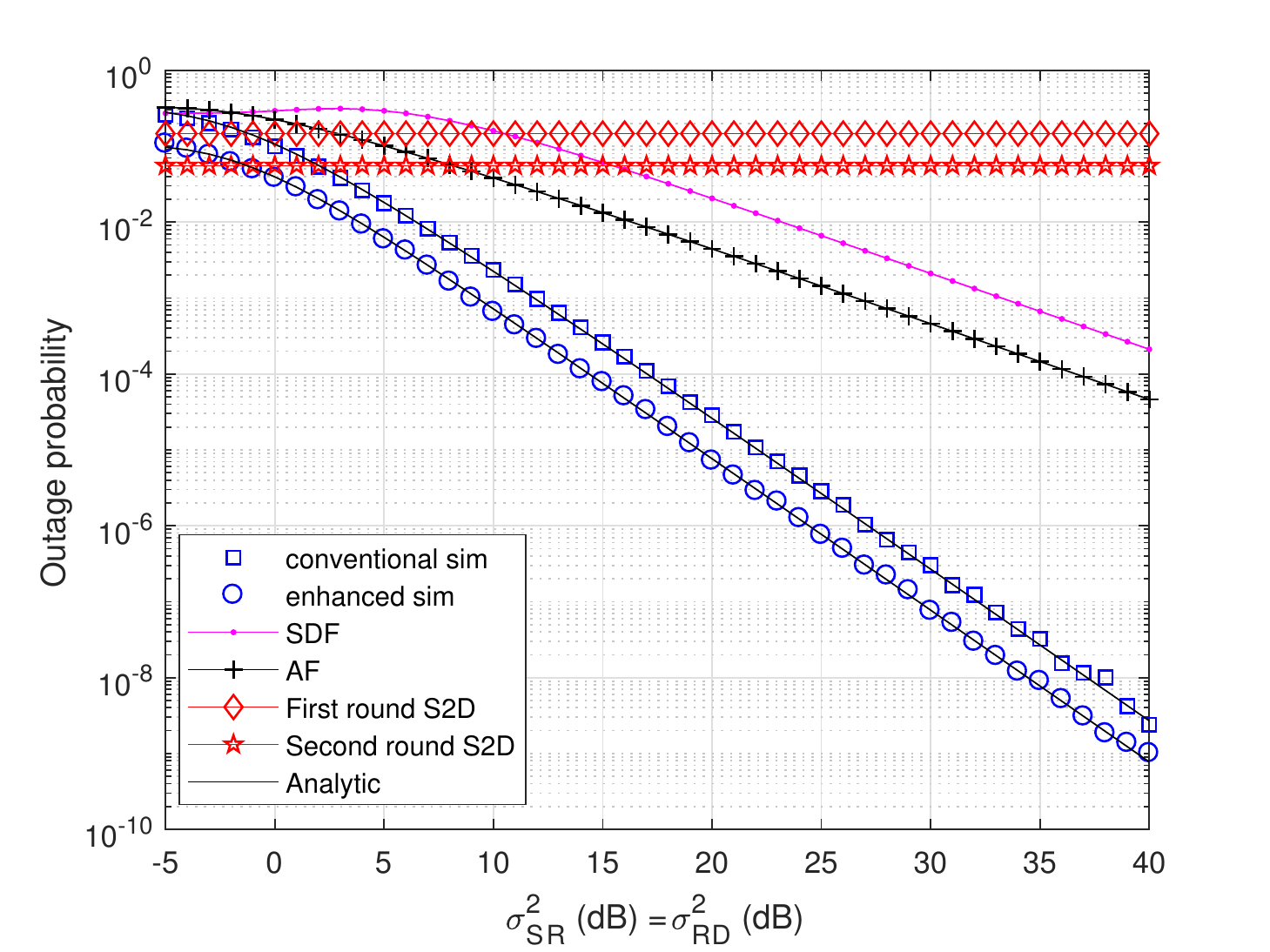}
\par\end{centering}

\centering{}\protect\caption{\label{fig:Outage-probabilitylow-versusSR-1-1} \small \textbf{The relaying link impact on the outage performance}. 
\\$\sigma_{\mathrm{SR}}^{2}=\sigma_{\mathrm{RD}}^{2}$,
$\sigma_{\mathrm{SD}}^{2}=0\,\mathrm{dB}$, $R=1$, and $\sigma_{\mathrm{RR}}^{2}=-10\,\mathrm{dB}$. }
\end{figure}

We further highlight the previously mentioned conclusions from different
perspectives in Fig. \ref{fig:Outage-probabilitylow-versusSD-1} and
Fig. \ref{fig:Outage-probabilitylow-versusSR-1-1}, where we investigate
the outage performance respectively, according to $\sigma_{\mathrm{SD}}^{2}$
and $\sigma_{\mathrm{SR}}^{2}$. Clearly Fig. \ref{fig:Outage-probabilitylow-versusSD-1}$.a$
points out the boosting effect of the direct link gain as it tends
to become gradually prominent. This favors mainly the use of a second
round transmission whether over a relay-based system, i.e., enhanced
HARQ or upon a S2D transmission. As can be seen, both of schemes have the
same slope with a large noted diversity gain regarding the first
round transmissions. However the proposed enhanced HARQ
procedure still the best in term of outage performance, mainly
due to the additional cooperative diversity brought through incorporating the $\mathrm{R}\rightarrow\mathrm{D}$
link. Indeed, with the increase of the direct link gain, the retransmission
opportunity results into more gains in terms of reliability. Specifically,
this gain is exhibited through the use of the direct link that interposes
to retransmit, whenever the relay is in outage. 

Regarding the nodes cooperation during the second round, Fig. \ref{fig:Outage-probabilitylow-versusSD-1}$.b$,
thanks to the good relaying links ($\sigma_{\mathrm{SR}}^{2}=\sigma_{\mathrm{RD}}^{2}=10\,\mathrm{dB}$),
the outage event on the relay-based system is less likely to happen
during the first round and thus the system nodes cooperation is reduced
during the second round. More specifically, at low $\mathrm{S\rightarrow D}$
link gains ($\sigma_{\mathrm{SD}}^{2}<0\,\textrm{dB}$), once a second
round transmission is required, the relay-based system will count
more on the relay involvement. On the other hand, at low direct link
gains, the outage probability of the S2D transmission is very high,
thereby the source cooperation becomes a necessity. Then, after the
sharp increase in the direct link gain, the system nodes cooperation
is extremely excluded. Nevertheless, the system nodes cooperation
in the relay-based system is still slightly higher than the S2D system
with a very negligible gain in term of outage probability. In summary,
for low $\mathrm{S\rightarrow D}$ link gains, the enhanced relay-based
HARQ system offers the best performance with low power consumption
during the second round. However, as the $\mathrm{S\rightarrow D}$
link becomes dominant, shutting down the relay seems to be the best
solution in term of both outage performance and power consumption.
For the rest of this section, we focus our analysis on low $\mathrm{S\rightarrow D}$
link case, i.e. $\sigma_{\mathrm{SD}}^{2}=0\,\textrm{dB}$. 

\begin{flushleft}
\begin{figure}[tbh]
\begin{centering}
\includegraphics[scale=0.50]{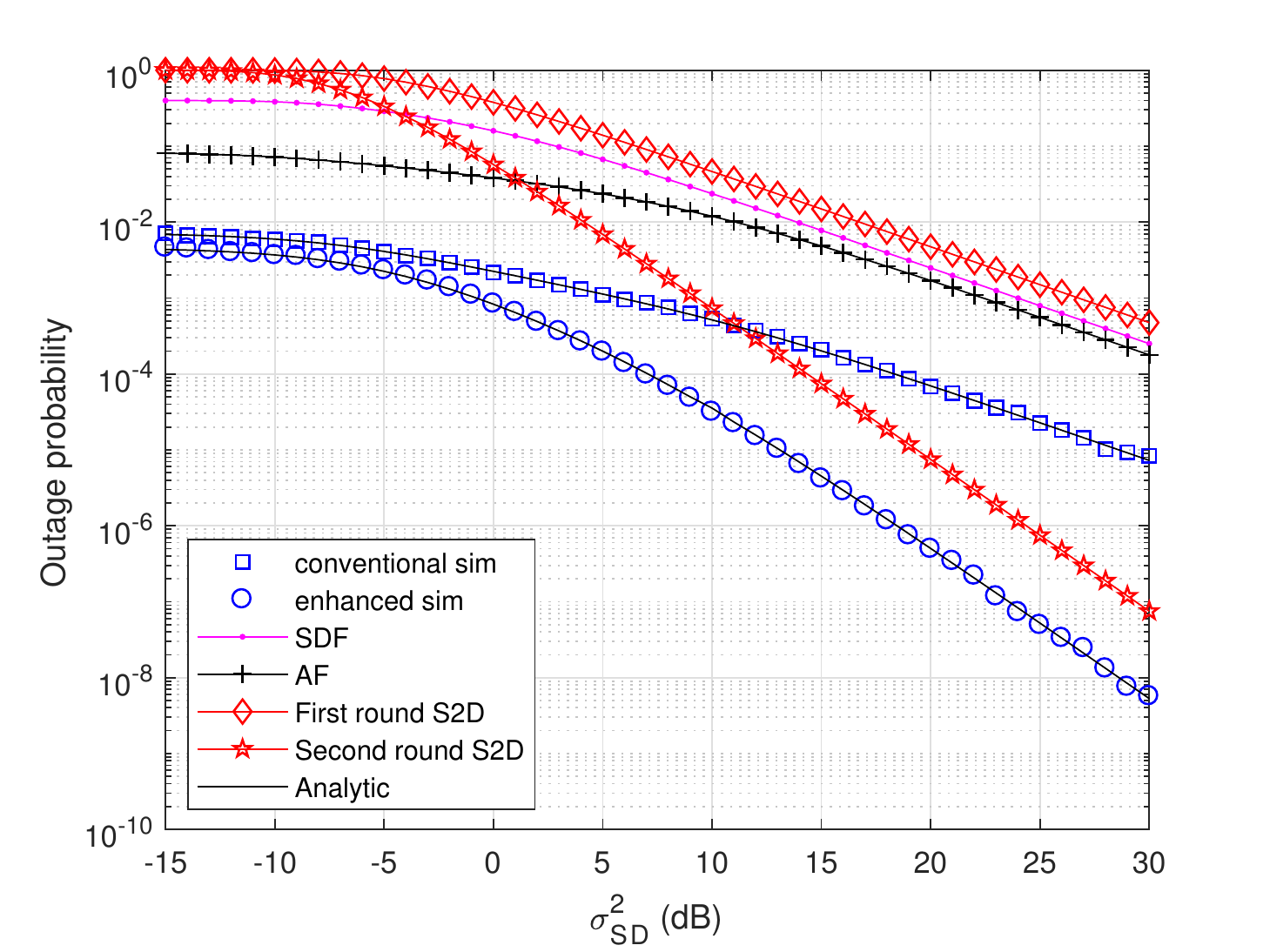}
\par\end{centering}

$\hphantom{\;}\hphantom{\;}\hphantom{\;}\hphantom{\;}\hphantom{\;}\hphantom{\;}\hphantom{\;}\hphantom{\;}\hphantom{\;}\hphantom{\;}\hphantom{\;}\hphantom{\;}\hphantom{\;}\hphantom{\;}\hphantom{\;}\hphantom{\;}\hphantom{\;}\hphantom{\;}\hphantom{\;}\hphantom{\;}\hphantom{\;}\hphantom{\;}\hphantom{\;} a)$
{\small{}Outage probability versus} $\sigma_{\mathrm{SD}}^{2}$.

\begin{centering}
\includegraphics[scale=0.50]{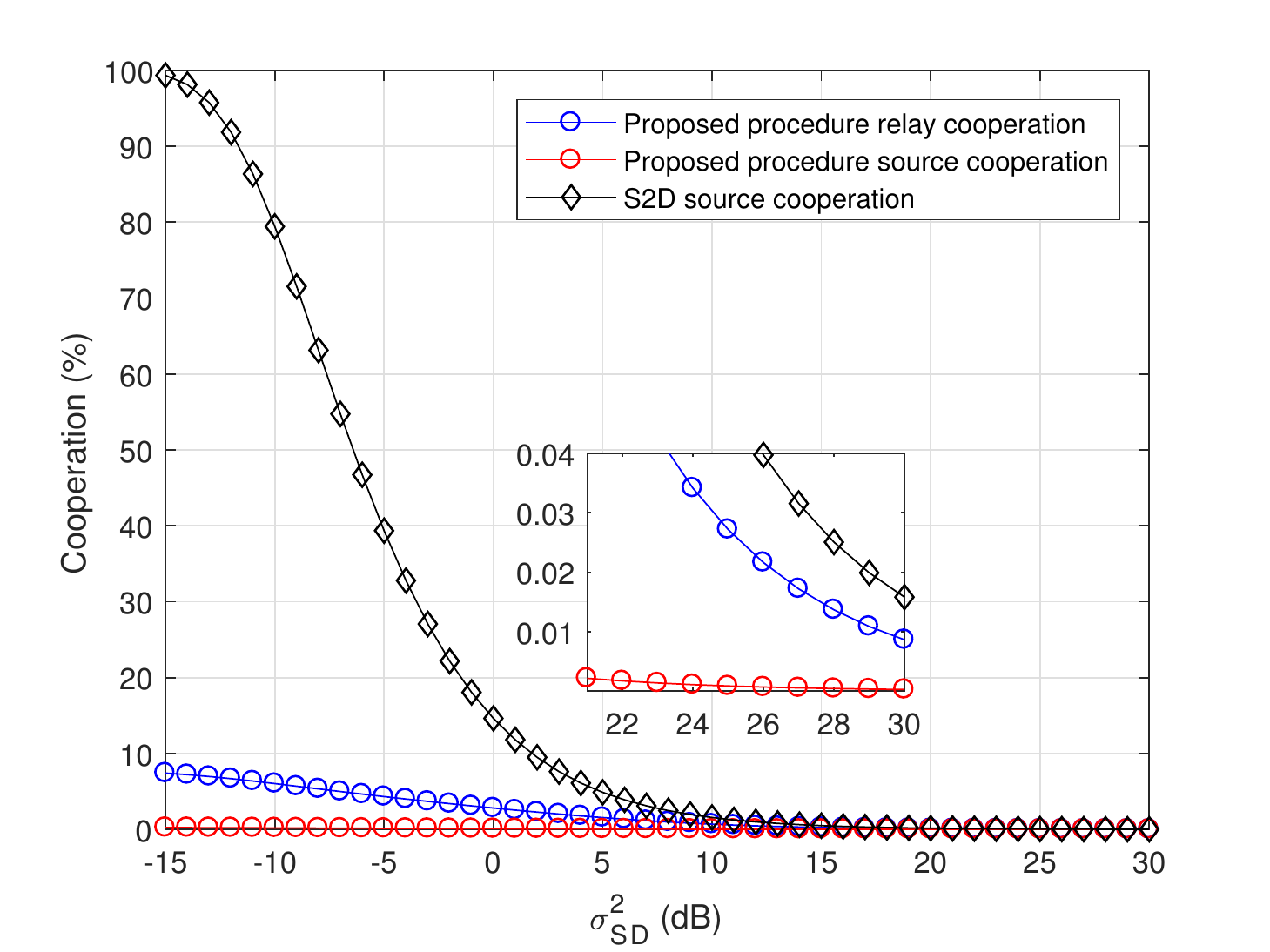}
\par\end{centering}
$\hphantom{\;}\hphantom{\;}\hphantom{\;}\hphantom{\;}\hphantom{\;}\hphantom{\;}\hphantom{\;}\hphantom{\;}\hphantom{\;}\hphantom{\;}\hphantom{\;}\hphantom{\;}\hphantom{\;}\hphantom{\;}\hphantom{\;}\hphantom{\;}\hphantom{\;}\hphantom{\;}\hphantom{\;}\hphantom{\;}\hphantom{\;}\hphantom{\;}\hphantom{\;} b)$ {\small{}Percentage
of retransmission.}{\small \par}

\centering{}\protect\caption{\label{fig:Outage-probabilitylow-versusSD-1} \small \textbf{The boosting impact of the direct link}. 
\\ $\sigma_{\mathrm{SR}}^{2}=\sigma_{\mathrm{RD}}^{2}=10\,\mathrm{dB}$,
$R=1$, and $\sigma_{\mathrm{RR}}^{2}=-10\,\mathrm{dB}$. }
\end{figure}

\par\end{flushleft}

Now, we consider in Fig. \ref{fig:Outage-probabilitylow-versusSR-1-1},
the scenario where the $\mathrm{S}\rightarrow\mathrm{R}$ and the
$\mathrm{R}\rightarrow\mathrm{D}$ link are varying equally, and the
direct link gain is brought to noise level, i.e., $\sigma_{\mathrm{SD}}^{2}=0\,\textrm{dB}$.
As noticed,  all cooperative schemes yield almost lower performance for a weak first hop link, where generally, the relay  is likely to be in outage, thereby no information will successfully reach the destination by the mean of the relay.
Otherwise, adopting a second round transmission for relay-based systems
(conventional and enhanced HARQ procedures) offers indeed better outage
performance as the $\mathrm{R}\rightarrow\mathrm{D}$ and $\mathrm{S}\rightarrow\mathrm{R}$
link gains become stronger. Note that whatever the first and second
hop link strengths, the enhanced HARQ procedure outperforms the conventional one however, with a reduced performance gap. This happens obviously because of the non apparent
use of the direct link that favors mainly the enhanced HARQ procedure.
On the other hand, for one round transmissions, still AF relaying
the best in term of outage performance.

\begin{figure}[h]
\centering{}\includegraphics[scale=0.50]{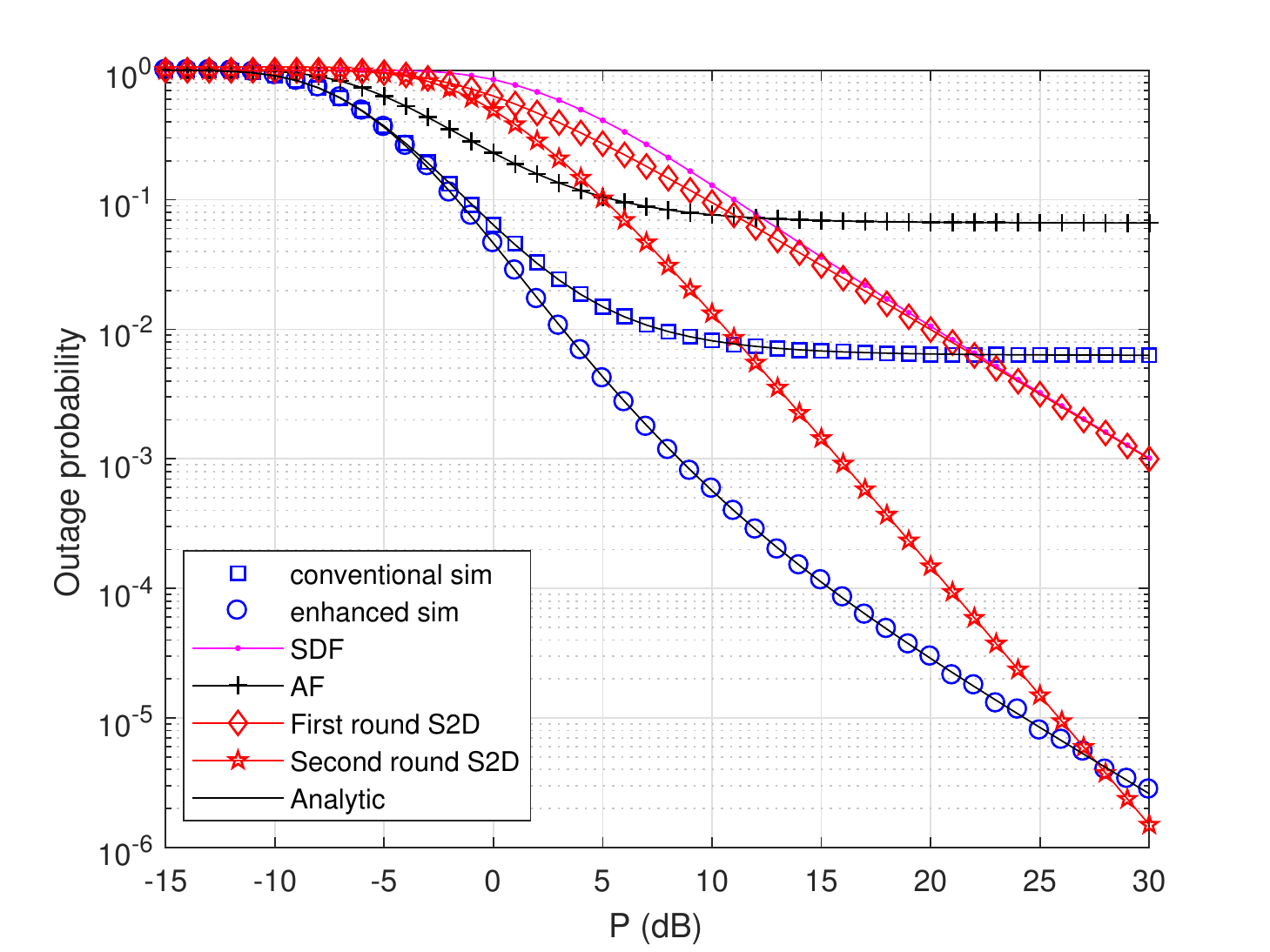}\protect\caption{\label{fig:Outage-probabilitylow-versusSNR_RSI0dB} \small \textbf{A worst-case scenario}. 
\\$\sigma_{\mathrm{SR}}^{2}=\sigma_{\mathrm{RD}}^{2}=10\,\mathrm{dB}$,
$\sigma_{\mathrm{SD}}^{2}=0\,\mathrm{dB}$, and $\sigma_{\mathrm{RR}}^{2}=0\,\mathrm{dB}$. }
\end{figure}

\begin{figure}[h]
\centering
\captionsetup{justification=centering}

\includegraphics[scale=0.50]{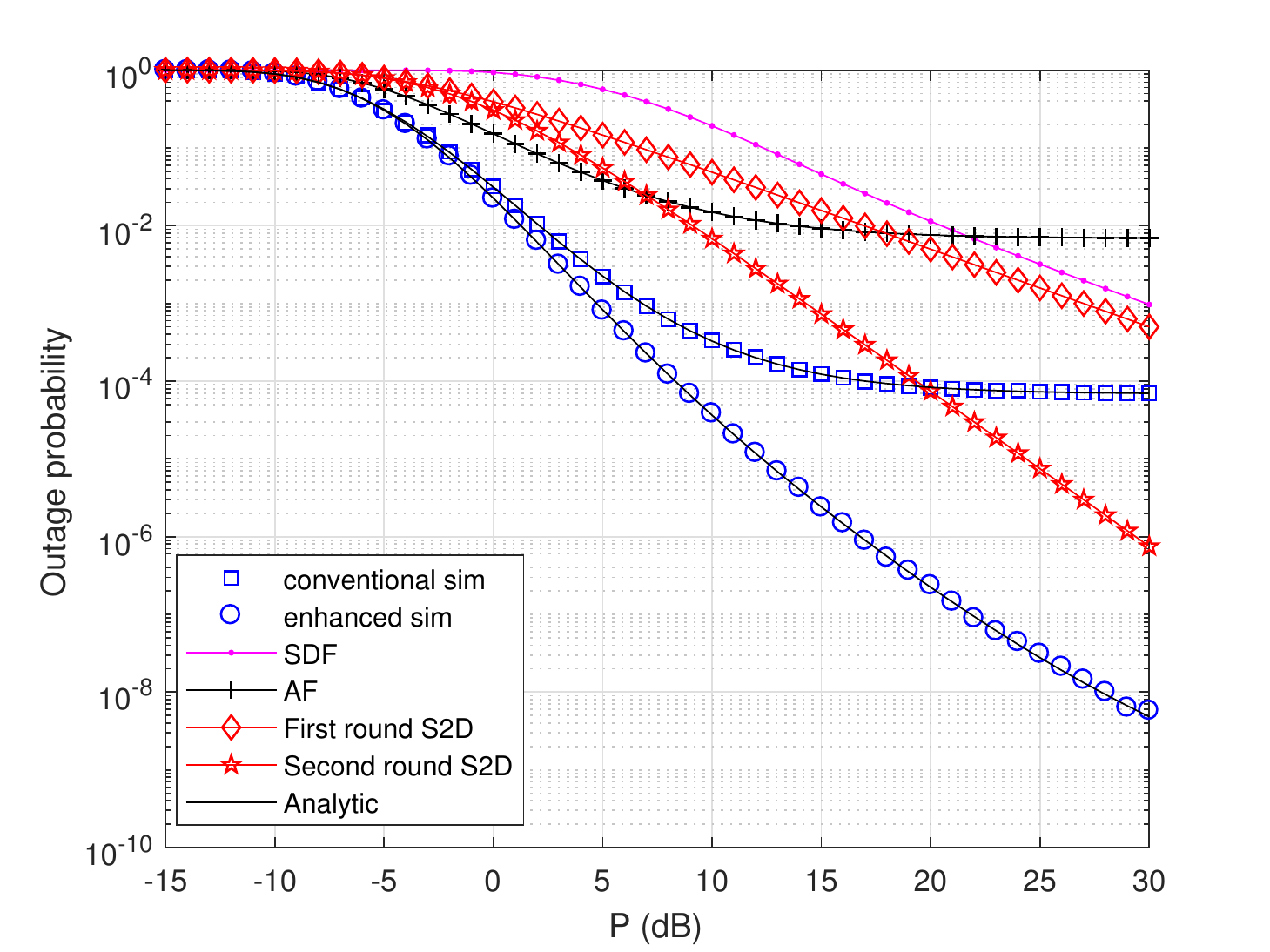}\protect

\caption{\label{fig:Outage-probabilitylow-versusSNR_RSI_minus10dB} \small \textbf{A realistic-case scenario}. 
\\$\sigma_{\mathrm{SR}}^{2}=\sigma_{\mathrm{RD}}^{2}=10\,\mathrm{dB}$,
$\sigma_{\mathrm{SD}}^{2}=0\,\mathrm{dB}$, and $\sigma_{\mathrm{RR}}^{2}=-10\,\mathrm{dB}$. } 
\end{figure}

To evaluate the diversity order of the different schemes, we plot,
in Fig. \ref{fig:Outage-probabilitylow-versusSNR_NoRSI}, Fig. \ref{fig:Outage-probabilitylow-versusSNR_RSI0dB},
and Fig. \ref{fig:Outage-probabilitylow-versusSNR_RSI_minus10dB},
the outage probability versus the transmit SNR level at the source
and the relay, both set to $P$. In Fig. \ref{fig:Outage-probabilitylow-versusSNR_NoRSI},
we ignore the RSI effect at the relay for all FD relaying schemes.
As shown, the first round S2D and the AF curves have the same slope
indicating a unity diversity order. As expected, using a second round
increases the diversity order for both direct and relay-based transmissions.
Using HARQ, the S2D transmission scheme and the conventional HARQ procedure reach the same classical diversity order of $2$. Indeed, despite the advantage of the good relaying links, the limited increase in the diversity order for the conventional procedure is
mainly due to the fact that when the $\mathrm{S\rightarrow R}$ is
in outage, the second round transmission is deactivated. On the other
hand, from Fig. \ref{fig:Outage-probabilitylow-versusSNR_NoRSI}, we notice that the enhanced HARQ procedure, where the source is involved during
the second round whenever the $\mathrm{S\rightarrow R}$ link is in
outage, enjoys of a much higher diversity order of $3$.
\begin{center}
\begin{figure}[h]
\centering{}\includegraphics[scale=0.50]{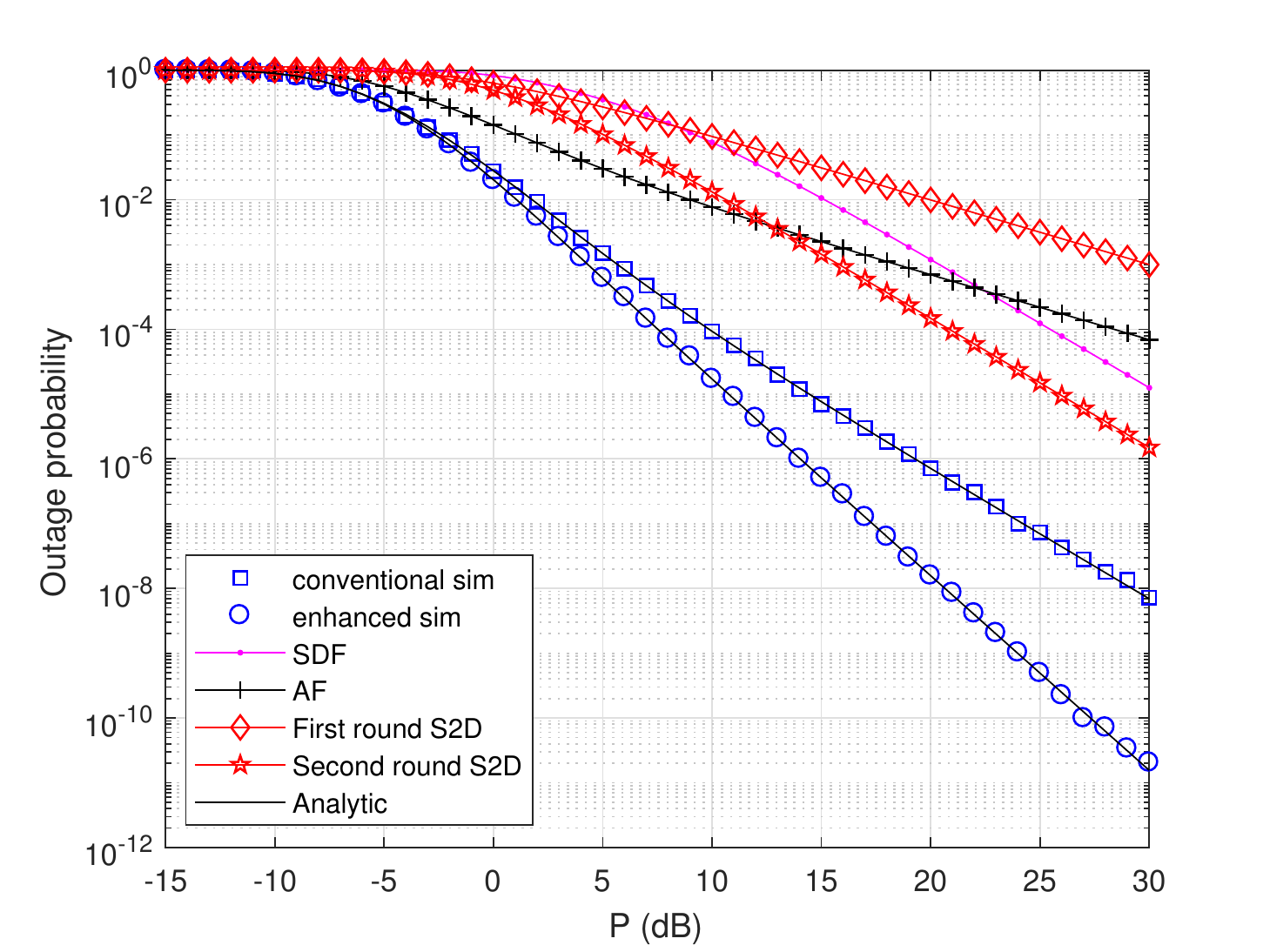}\protect\caption{\label{fig:Outage-probabilitylow-versusSNR_NoRSI} \small \textbf{The perfect scenario}. 
\\$\sigma_{\mathrm{SR}}^{2}=\sigma_{\mathrm{RD}}^{2}=10\,\mathrm{dB}$,
$\sigma_{\mathrm{SD}}^{2}=0\,\mathrm{dB}$, and negligible RSI. }
\end{figure}

\par\end{center}

In Fig. \ref{fig:Outage-probabilitylow-versusSNR_RSI0dB}, we consider
the case where the RSI gain in $\gamma_{\mathrm{SR}}=\frac{P_{\textrm{S}}|h_{\mathrm{\mathrm{SR}}}|^{2}}{P_{\mathrm{R}}\sigma_{\mathrm{RR}}^{2}+N_{R}}$
scales linearly with the relay transmit power, i.e., $\sigma_{\mathrm{RR}}^{2}=0\,\mathrm{dB}$.
We notice that all FD-relaying schemes diversity are badly affected
by the loop-back interference at the relay side, i.e., the conventional
HARQ procedure experiences an error floor while the enhanced HARQ
procedure diversity order drops to $1$. Still the proposed schemes
offer the best performance for moderate transmit power values.
However, as shown in the Fig. \ref{fig:Outage-probabilitylow-versusSNR_RSI_minus10dB},
the performance of the FD-relaying schemes can be significantly enhanced if
the adopted loopback interference isolation and cancellation techniques
can dramatically reduce the RSI effect, i.e., $\sigma_{\mathrm{RR}}^{2}=-10\,\mathrm{dB}$.

\section{\label{sec:Conclusion}Conclusion}

In this paper, we investigated a hybrid FD AF-SDF relay-based system
and we proposed two retransmission procedures within which the HARQ
RTT is shortened as a path towards URLLC. Once a packet is received
in error, the studied system might move along one of suggested retransmission
procedures, namely, the conventional and the enhanced  HARQ procedures.
Thereby, to capture their inherent performance difference, we derived the
analytical outage probability expression
respectively, for the conventional and the enhanced mechanisms. Then
we compared them to the most known relaying techniques, i.e., AF as
the low latency relay processing scheme, and the SDF scheme that requires
complex encoding and decoding algorithms. Furthermore for a fair comparison,
we included the non-cooperative mode, where the direct link handles
both, the transmission and the retransmission whenever a second round
transmission becomes a necessity. Indeed, both of procedures allow
a prompt processing, however as simulation results have revealed,
the enhanced one offers additional improvements in term of reliability
and eventually further available diversity via the direct link whenever
the RSI is negligible.

\end{document}